\RequirePackage{tikz,pgf}
\RequirePackage{tikz-qtree,tikz-qtree-compat}

\documentclass[11pt,final]{amsart}       % one column (second format)

 \usepackage{epsfig, epsf, float, color}
 \usepackage{amssymb,amsaddr,paralist}
 \usepackage{verbatim,hyperref}
 \usepackage{setspace,mathtools,wrapfig}
 \usepackage{steinmetz}
 \usepackage{arydshln}
 \usepackage{cleveref}

\usepackage[mathlines]{lineno}

 \usetikzlibrary{decorations,decorations.markings,decorations.text}
 \usetikzlibrary{arrows.meta}
 \usetikzlibrary{patterns}
 \usetikzlibrary{positioning,patterns}
 \usetikzlibrary{calc,fadings,decorations.pathreplacing,arrows,datavisualization.formats.functions,shapes.geometric}

 \usepackage{lipsum}
 \usepackage{amsfonts}
 \usepackage{graphicx}
 \usepackage{epstopdf}
 \usepackage{algorithm}
 \usepackage{algpseudocode}

\pgfkeys{/pgf/decoration/.cd,
      distance/.initial=10pt
}  

\pgfdeclaredecoration{add dim}{final}{
\state{final}{\pgfmathsetmacro{\dist}{5pt*\pgfkeysvalueof{/pgf/decoration/distance}/abs(\pgfkeysvalueof{/pgf/decoration/distance})}    
          \pgfpathmoveto{\pgfpoint{0pt}{0pt}}             
          \pgfpathlineto{\pgfpoint{0pt}{2*\dist}}   
          \pgfpathmoveto{\pgfpoint{\pgfdecoratedpathlength}{0pt}} 
          \pgfpathlineto{\pgfpoint{(\pgfdecoratedpathlength}{2*\dist}}     
          \pgfsetarrowsstart{latex}
          \pgfsetarrowsend{latex}
          \pgfpathmoveto{\pgfpoint{0pt}{\dist}}
          \pgfpathlineto{\pgfpoint{\pgfdecoratedpathlength}{\dist}} 
          \pgfusepath{stroke} 
          \pgfpathmoveto{\pgfpoint{0pt}{0pt}}
          \pgfpathlineto{\pgfpoint{\pgfdecoratedpathlength}{0pt}}
}}

\tikzset{dim/.style args={#1,#2}{decoration={add dim,distance=#2},
                decorate,
                postaction={decorate,decoration={text along path,
                                                 raise=#2,
                                                 text align={align=center},
                                                 text={#1}}}}}

\clearpage{}

\def\boxit#1{\vbox{\hrule\hbox{\vrule\kern6pt
          \vbox{\kern6pt#1\kern6pt}\kern6pt\vrule}\hrule}}

\def\cov{\hbox{cov}}

\def\bse{\begin{eqnarray*}}
\def\ese{\end{eqnarray*}}
\def\be{\begin{eqnarray}}
\def\ee{\end{eqnarray}}
\def\bq{\begin{equation}}
\def\eq{\end{equation}}
\def\bse{\begin{eqnarray*}}
\def\ese{\end{eqnarray*}}

\def\T{^{\rm T}}

\newcommand{\corb}[1]{\textcolor{black}{#1}}

\newcommand{\bD}{\mathbf{D}}
\newcommand{\bI}{\mathbf{I}}
\newcommand{\bL}{\mathbf{L}}
\newcommand{\bG}{\mathbf{G}}
\newcommand{\bW}{\mathbf{W}}
\newcommand{\bP}{\mathbf{P}}
\newcommand{\bU}{\mathbf{U}}
\newcommand{\bC}{\mathbf{C}}
\newcommand{\bA}{\mathbf{A}}

\newcommand{\bS}{\mathbf{S}}
\newcommand{\bV}{\mathbf{V}}
\newcommand{\bX}{\mathbf{X}}

\newcommand{\bZ}{\mathbf{Z}}

\newcommand{\bx}{\mathbf{x}}
\newcommand{\by}{\mathbf{y}}
\newcommand{\bv}{\mathbf{v}}
\newcommand{\bz}{\mathbf{z}}

\newcommand{\ba}{\mathbf{a}}
\newcommand{\bb}{\mathbf{b}}

\newcommand{\bc}{\mathbf{c}}

\newcommand{\bve}{\mathbf{e}}
\newcommand{\bu}{\mathbf{u}}
\newcommand{\bw}{\mathbf{w}}
\newcommand{\bm}{\mathbf{m}}
\newcommand{\bg}{\mathbf{g}}

\newcommand{\bk}{\mathbf{k}}

\newcommand{\bgamma}{\boldsymbol{\gamma}}

\newcommand{\blambda}{\boldsymbol{\lambda}}
\newcommand{\bdelta}{\boldsymbol{\delta}}
\newcommand{\btheta}{\boldsymbol{\theta}}

\newcommand{\bpsi}{\boldsymbol{\psi}}
\newcommand{\bphi}{\boldsymbol{\phi}}
\newcommand{\bepsilon}{\boldsymbol{\epsilon}}
\newcommand{\balpha}{\boldsymbol{\alpha}}
\newcommand{\bbeta}{\boldsymbol{\beta}}

\newcommand{\0}{\mathbf{0}}

\newcommand{\mcB}{{\mathcal B}}
\newcommand{\mcC}{{\mathcal C}}
\newcommand{\mcD}{{\mathcal D}}
\newcommand{\mcE}{{\mathcal E}}

\newcommand{\mcG}{\mathcal{G}}
\newcommand{\mcI}{{\mathcal I}}

\newcommand{\mcO}{{\mathcal O}}
\newcommand{\mcN}{{\mathcal N}}
\newcommand{\mcM}{{\mathcal M}}
\newcommand{\mcP}{{\mathcal P}}
\newcommand{\mcQ}{{\mathcal Q}}

\newcommand{\mcR}{{\mathcal R}}
\newcommand{\mcS}{\mathcal{S}}

\newcommand{\mcX}{\mathcal{X}}
\newcommand{\mcY}{\mathcal{Y}}

\newcommand{\bbR}{\mathbb{R}}

\newcommand{\bbN}{\mathbb{N}}
\newcommand{\bbE}{\mathbb{E}}

\newcommand{\bbS}{\mathbb{S}}
\newcommand{\bbK}{\mathbb{K}}
\newcommand{\bbP}{\mathbb{P}}
\newcommand{\bbC}{\mathbb{C}}

\newcommand{\Nset}{\mathbb{N}_0}

\newcommand{\bbNset}{{\mathbb N}}

\newcommand{\dist}{\operatorname{dist}}

\newcommand{\Real}{\mathop{\text{\rm Re}}}
\newcommand{\Imag}{\mathop{\text{\rm Im}}}

\def\R{\Bbb{R}}

\newcommand{\argmax}{\operatornamewithlimits{argmax}}

\def\boxit#1{\smash{\color{blue}\fboxrule=1pt\relax\fboxsep=2pt\relax \llap{\rlap{\fbox{\vphantom{0}\makebox[#1]{}}}~}}\ignorespaces
}

\definecolor{darkblue}{rgb}{0,0.08,0.4}
\definecolor{brightblue}{rgb}{0.65,0.85,0.85}
\definecolor{darkred}{rgb}{0.8,0.2, 0.2}
\definecolor{darkgreen}{rgb}{0, 0.6, 0}
\definecolor{blueish}{rgb}{0.1176, 0.5647, 1.0000}
\definecolor{babyblue}{RGB}{153,204,255}
\definecolor{darkorange}{RGB}{255,140,0}

\definecolor{colorone}{rgb}{0.1176,0.5647,1.0000}
\definecolor{colortwo}{rgb}{0.5608,0.7373,0.5608}
\clearpage{}

\theoremstyle{thmstyleone}\newtheorem{theorem}{Theorem}\newtheorem{proposition}[theorem]{Proposition}

\newtheorem{lemma}[theorem]{Lemma}

\theoremstyle{thmstyletwo}\newtheorem{remark}{Remark}
\theoremstyle{thmstylethree}\newtheorem{definition}{Definition}\newtheorem{asum}{Assumption}
\begin{document}

\title[Spatial best linear unbiased prediction for massive datasets]
{
  Spatial best linear unbiased prediction:
    A computational 
    mathematics approach for high dimensional massive datasets}

\author{Julio E. Castrill\'on-Cand\'as ${\dagger}$} 
  \email{jcandas@bu.edu}

 \address{
   ${\ddagger}$ Department of Mathematics and Statistics, 
  Boston University, Boston, MA 
  }

 \begin{abstract}
With the advent of massive data sets much of the computational science
and engineering community has moved toward data-intensive approaches
in regression and classification. However, these present significant
challenges due to increasing size, complexity and dimensionality of
the problems. In particular, covariance matrices in many cases are
numerically unstable and linear algebra shows that often such matrices
cannot be inverted accurately on a finite precision computer.  A
common ad hoc approach to stabilizing a matrix is application of a
so-called nugget. However, this can change the model and introduce
error to the original solution. \emph{It is well known from numerical
analysis that ill-conditioned matrices cannot be accurately inverted.}
In this paper we develop a multilevel computational method that scales
well with the number of observations and dimensions.  A multilevel
basis is constructed adapted to a kD-tree partitioning of the
observations.  Numerically unstable covariance matrices with large
condition numbers can be transformed into well conditioned multilevel
ones without compromising accuracy. Moreover, it is shown that the
multilevel prediction \emph{exactly} solves the Best Linear Unbiased
Predictor (BLUP) and Generalized Least Squares (GLS) model, but is
numerically stable.  The multilevel method is tested on numerically
unstable problems of up to 25 dimensions. Numerical results show
speedups of up to 42,050 times for solving the BLUP problem, but with
the same accuracy as the traditional iterative approach. For very
ill-conditioned cases the speedup is infinite.  In addition, decay
estimates of the multilevel covariance matrices are derived based on
high dimensional interpolation techniques from the field of numerical
analysis.  This work lies at the intersection of statistics,
uncertainty quantification, high performance computing and
computational applied mathematics.
\end{abstract}

\maketitle



\section{Introduction}

Massive data sets arise from many fields, including, but not limited
to commerce, astrophysical sky-surveys, environmental data, medical
data, and tsunami warning systems. With the advent of massive
datasets, much of the computational science and engineering community
has moved toward
data-intensive approaches in regression and classification. However,
these present significant challenges due to increasing size,
complexity, and dimensionality of the problems.

Best Linear Unbiased Prediction (BLUP), or sometimes referred also the
best linear unbiased predictor is a well known technique in earth and
environmental sciences \cite{Cressie2006, Cressie1993}. It was
originally developed by Henderson \cite{Henderson1950, Henderson1975}
in the context of biosciences and biostatistics. It is also popular in
Longitudinal analysis. This field is very significant to gerontology
as well as the biomedical, and behavioral and social sciences
\cite{Liu2016, Liu2008}. \corb{Stein's book, referenced as
  \cite{Stein1999}, provides valuable insights into the topic of
  BLUPs.}

Since solving for the BLUP requires inverting the covariance matrix,
this in general requires $\mcO(N^3)$ computational steps and
$\mcO(N^2)$ memory \cite{Cressie2006}. For massive datasets this
quickly becomes intractable since: (I) The covariance matrix becomes
too large, and (II) for spatial covariance functions, the problem is
further compounded by ill-conditioning of the covariance
matrix. \emph{It is known from linear algebra that ill-conditioned
matrices cannot be accurately inverted with accuracy on finite
precision computers \cite{Golub96} and thus are difficult, if not
impossible, to solve numerically}.

A common technique to correct the ill-conditioned covariance matrix
$\bC$ is to add a scaled identity matrix $\bI$, e.g. $\bC + \sigma
\bI$, where $\sigma > 0$. The term $\sigma \bI$ is a called a
\emph{nugget}. However, inverting the matrix $\bC + \sigma \bI$ is not
equivalent to inverting $\bC$. \corb{The solution of the BLUP will be
incorrect.} Thus a tradeoff between accuracy and numerical
stability is commonly accepted since if a matrix is ill-conditioned, it
cannot be accurately inverted.

Many techniques for inverting $\bC$ concentrate on the first problem
(I). They rely on sparsification and/or identifying low rank
approximations.  In the context of estimating the covariance function,
many methods have been developed using skeletonization factorizations
\cite{Minden2016}, low-rank \cite{nowak2013} and Hierarchical Matrices
(HM) \cite{khoromskij2009,Litvinenko2019,Geoga2020} approaches. These
methods are very promising. In particular, for the HM approaches they
have been shown to be near optimal. They work well for low dimensions.
However, as the dimensions increases the computational burden explodes
with each dimension. However, they are still subject to
ill-conditioning and usually a nugget is added to change the model to
make it more numerically stable, but does not solve the original
problem. \corb{Thus these approaches are limited to covariance
  matrices that are well conditioned.}  Moreover, the model of the
data is assumed to have zero trend and a non-zero nugget. For many
practical cases this will not be valid. \corb{Note that in
  \cite{Schafer2020,Schafer2021} the authors developed an approach for
  constructing sparse positive definite kernel matrices with improved
  numerical stability.}

\corb{In the approach developed in \cite{Castrillon2015} it is shown
  that there exists a more stable form of the solution of the BLUP and
  the Generalized Least Square (GLS) that solves these problems
  exactly. This approach is based on the work on multilevel discrete
  basis developed in \cite{Castrillon2013} for the radial basis
  function interpolation problem for scattered data. A similar basis
  has been developed in \cite{Harbrecht2022}. These approaches are
  based on the idea of using wavelets to compress integral operators
  \cite{Beylkin1991}.}

\corb{Although an ill-conditioned covariance matrix leads to accuracy
  problems, this can be avoided by constructing an alternative
  multilevel covariance matrix $\bC_{\bW}$ that is used in the stable
  form.} It is also shown how the covariance matrix can be sparsified
so that the covariance function can be estimated from the data using a
Maximum Likelihood Estimate method.  Error estimates for the decay of
the covariance function are derived using derivative information of
the covariance function.  However, this approach is limited to 2 or 3
dimensions. The computational cost scales combinatorially fast with
the spatial dimension, thus making it impractical for high dimensional
problems.

In this paper the approach from \cite{Castrillon2015} is extended
using binary trees, which are well suited for high dimensional
problems. \corb{The multilevel basis used in \cite{Castrillon2015} and
  originally proposed in \cite{Castrillon2013} is extended to the high
  dimensional setting.}  Ill-conditioned covariance matrices are
transformed to numerically stable multilevel covariance matrices
without compromising accuracy. In addition, a new distance criterion
is developed to build sparse multilevel covariance matrices.
Furthermore, sharper decay estimates of the coefficients of the
multivariate covariance matrix are derived based on analytic
extensions that are well suited for high dimensional problems.

\corb{In the research presented in \cite{Castrillon2015}, the authors
  establish the decay rates of covariance matrix entries for
  multilevel matrices in $\mathbb{R}^{d}$ using Taylor's
  theorem. While this method is effective for lower-dimensional
  problems, its practicality diminishes as the dimensionality, denoted
  as $d$, increases. This is due to the combinatorial growth in the
  number of required derivatives in Taylor's theorem, alongside the
  expansion of the derivative domain concerning the dimension $d$. In
  cases involving high-dimensional scenarios, the computation of
  constants related to the derivatives in the results of
  \cite{Castrillon2015} becomes increasingly challenging. In contrast,
  the complex analytic approach offers the advantage of uniformly
  bounding these constants, which depend on the region of analytic
  extension. As a result, in the field of uncertainty quantification
  for stochastic Partial Differential Equations featuring
  high-dimensional random parameters, complex analyticity is favored
  \cite{nobile2008a,nobile2008b,Castrillon2016,Castrillon2021a,Castrillon2021b}. This
  approach is adopted in this paper.}

The MLE estimation equations are transformed into a multilevel form
based on the numerically stable multilevel covariance matrix. In
practice a sparse version of the multilevel covariance matrix is
used. A distance dependent method is used to build to a sparse
version. Sharp decay estimates (sub-exponential) of the multilevel
covariance matrices are derived using complex analytic extensions of
the covariance function instead of Taylor series expansions, which are
infeasible for relatively large dimensional problems. The numerical
results show that the estimation is solved to good accuracy for a
large number of observations.

The BLUP prediction step is remapped into an equivalent multilevel
formulation that is numerically stable. \corb{It is shown that the
  solution to the multilevel prediction form \emph{exactly} solves the
  BLUP problem.}  To my knowledge, this is a feature that is unique to
the multilevel approach.  If the covariance matrix $\bC$ is
ill-conditioned, then it is not possible to solve the problem
accurately on a computer with a fixed machine precision. However, the
BLUP solution arises from a constrained optimization problem. \corb{By
taking advantage of this fact, the multilevel approach side steps the
inversion of the covariance matrix and directly searches for the
solution in a constrained space giving rise to a significantly more
stable multilevel covariance matrix.}  \corb{Moreover, by using an
  iterative approach only one indirect matrix inversion of the
  multilevel covariance $\bC_{\bW}$ matrix is required. This is in
  contrast to classical BLUP, including the Generalized Least Squares
  (GLS) prediction, that at least $p$ indirect matrix inversions are
  required with an iterative approach, where $p$ is the number of
  columns of the design matrix (See Remark \ref{Introduction:remark1}
  and \ref{Spatial:remark1}).}  Numerical results show speedups of up
to 42,050 for solving the BLUP problem to at least the same
accuracy. \corb{This approach has been also applied for imputation of
  medical records \cite{Li2023}}.

In Section \ref{Introduction} the problem formulation is introduced.
In section \ref{multilevelapproach} it is shown how to construct the
multilevel basis based on kd-trees.  In section
\ref{MultilevelCovarianceMatrix} the construction of the multilevel
covariance matrix is discussed.  In section \ref{multilevelestimator}
the multilevel estimator and predictor are formulated and numerical
computational issues are discussed in section
\ref{numericalcomputation}.  In section \ref{errorestimates} a
mathematical analysis of the decay of the entries of the multilevel
covariance matrix is developed. This section can also be skipped for
the less mathematically inclined reader.  In section
\ref{numericalresults} the multilevel BLUP method is tested on
numerically unstable problems of up to 25 dimensions. Furthermore, a
direct accuracy comparison is done with the traditional BLUP formulae.
Highly ill-conditioned BLUP problems are solved to high accuracy.  In
the appendices all of the proofs are described in detail and in
Appendix \ref{PolynomialAppendix} a the multivariate polynomial
interpolation based on complex analytic extensions is discussed. These
results are used for to derive the decay of the entries of the
multilevel covariance matrix.

\section{Problem setup}
\label{Introduction}

Consider the following model for a Gaussian random field $Z$:
\begin{equation}
Z(\bx) = \bk(\bx)\T \bbeta+\varepsilon(\bx), \qquad \bx \in \R^d,
\label{Introduction:noisemodel}
\end{equation}
where $d$ is the number of spatial dimensions, $\bk:\R^d \rightarrow
\R^p$ is a functional vector of the spatial location $\bx$,
$\bbeta\in\R^p$ is an unknown vector of coefficients, and
$\varepsilon$ is a stationary mean zero Gaussian random field with
parametric covariance function
\corb{$\phi(\bx,\bx';\btheta)=\cov\{\varepsilon(\bx),\varepsilon(\bx')\}$
  with an unknown vector of positive parameters $\btheta\in\R^w$,
  where $w$ is the number of parameters.}

Suppose that we obtain $N$ observations and stack them in the data
vector $\bZ=(Z(\bx_1),\ldots,$ $Z(\bx_N))\T$ from locations $\bbS :=\{
\bx_{1},\dots,\bx_{N}\}$, where the elements in $\bbS$ are restricted
such that the design matrix defined below, $\bX$, has full column
rank.  \corb{Furthermore, without loss of generality all the locations
  in $\bbS$ are contained in the unit hypercube $\Gamma^{d} \equiv
  [-1,1]^{d}$.}  Let $\bC(\btheta)=\cov(\bZ,\bZ\T)\in \R^{N \times N}$
be the covariance matrix of $\bZ$ and assume it is positive definite
for all $\btheta\in\R^w$.  Define $\bX=\big( \bk(\bx_1) \ldots$ $
\bk(\bx_N)\big)\T\in \R^{N\times p}$ and assume it is of full rank
$p$. Since the model \eqref{Introduction:noisemodel} is a Gaussian
random field, then from the samples of $\bbS$ the following vectorial
model is obtained
\begin{equation}
\bZ = \bX \bbeta +{\boldsymbol \varepsilon},
\label{Introduction:vectormodel}
\end{equation}
\corb{where $\boldsymbol \varepsilon$ is a Gaussian random vector,
${\boldsymbol \varepsilon} \sim \mcN(\0,\bC(\btheta))$ and $p < N$}.
The aim now is to: i) {\it Estimate} the unknown vectors $\bbeta$ and
$\btheta$; and ii) {\it Predict} $Z(\bx_0)$, where $\bx_0$ is a new
spatial location. These two tasks are particularly computationally
challenging when the sample size $N$ and number of dimensions $d$ are
large.

There is a very large literature on Gaussian process regression that
deal with this problem. Please see \cite{Castrillon2015} for a brief
literature review.  The unknown vectors $\bbeta$ and $\btheta$ are
estimated with the log-likelihood function $\ell(\bbeta,\btheta)
=-\frac{n}{2}\log(2\pi)-\frac{1}{2}\log \det\{\bC(\btheta)\}
-\frac{1}{2}(\bZ-\bX\bbeta)\T\bC(\btheta)^{-1}
(\bZ-\bX\bbeta)$. \corb{To reduce the dimensionality of the
  optimization problem, $\bbeta$ is replaced with GLS estimate:}
\begin{equation}
  \hat \bbeta(\btheta)=\{\bX\T \bC(\btheta)^{-1} \bX\}^{-1}\bX\T\bC(\btheta)^{-1}\bZ.
  \label{GLSbeta}
\end{equation}
In general this is not a good choice, since replacing with the Maximum
Likelihood Estimator (MLE) of $\btheta$ is prone to be biased
\cite{Castrillon2015}.

For the prediction part, consider the BLUP $\hat
Z(\bx_0)=\lambda_0+\blambda\T\bZ$ where
$\blambda=(\lambda_1,\ldots,\lambda_N)\T$. The unbiased constraint
implies $\lambda_0=0$ and $\bX\T\blambda=\bk(\bx_0)$.  The
minimization of the mean squared prediction error
E$[\{Z(\bx_0)-\blambda\T\bZ\}^2]$ under the \corb{unbiased} constraint
$\bX\T\blambda=\bk(\bx_0)$ yields
\begin{equation}
\hat Z(\bx_0)=\bk(\bx_0)\T\hat \bbeta+\bc(\btheta)\T
\bC(\btheta)^{-1}(\bZ-\bX\hat \bbeta), \label{KrigBLUP}
\end{equation}
where $\bc(\btheta)=\cov\{\bZ,Z(\bx_0)\}\in \R^{N}$ and $\hat \bbeta$ is
defined in (\ref{GLSbeta}).

\begin{remark}
  \corb{Notice that solving for $\hat \bbeta(\btheta)$ requires
    computing $\bC(\btheta)^{-1} \bX$. Since $\bX \in \R^{N \times p}$
    by using an iterative approach it would require $p$ indirect
    inversions of the matrix $\bC(\btheta)^{-1}$.}
  \label{Introduction:remark1}
  \end{remark}

Now, let $\alpha:= (\alpha_{1},\dots,\alpha_{d}) \in \mathbb{Z}{^d}$,
$\lvert \alpha \rvert := \alpha_{1}+\dots+\alpha_{d}$, $\bx : =
[x_1,\dots,x_d]$. For any $w \in \bbN_+$ (where $\mathbb{N}_+ :=
\mathbb{N} \cup \{0\}$) let $\mcQ^d_w$ be the set of Total Degree (TD)
monomials $\{x_1^{\alpha_1} \dots x_d^{\alpha_d}\,\,\,\vert\,\,\, \lvert \alpha \rvert
\leq w\}$. The typical choice for the matrix $\bX$ is to build it from
the monomials of $\mcQ^d_w$ with cardinality
$p(d,w):=\begin{pmatrix} d + w \\ w \end{pmatrix}$.

The challenge is that the covariance matrix $\bC(\btheta)$ in many
practical cases is ill-conditioned, leading to slow and inaccurate
estimates of $\btheta$. Following the approach in
\cite{Castrillon2015} the data vector $\bZ$ is transformed into
decoupled multilevel description of the model
\eqref{Introduction:noisemodel}.  This multilevel representation leads
to significant computational benefits, including numerical stability,
when computing the multilevel predictor $\hat Z(\bx_0)$ in
(\ref{KrigBLUP}) for large sample size $N$ and high dimensions $d$.
Note, that in this paper we shall refer to the \emph{single level}
approach to solving the estimation and prediction steps directly to
the data $\bZ$ and covariance matrix $\bC(\btheta)$.

\section{Multilevel approach}
\label{multilevelapproach}

The general approach of this paper and multilevel basis construction
are now presented. We mostly follow the exposition laid out in
\cite{Castrillon2015}. The proof of Proposition
\ref{Multilevelapproach:theo1} is repeated, but clarified with
more details.

Let $\mcP^{p}(\bbS)$ be the span of the columns of the design matrix
$\bX$. Suppose that there exists the orthogonal projections $\bL :
\R^N \rightarrow \mcP^{p}(\bbS)$ and $\bW : \R^N \rightarrow
\mcP^{p}(\bbS)^{\perp}$, where $\mcP^{p}(\bbS)^{\perp}$ is the
orthogonal complement of $\mcP^{p}(\bbS)$.  The operator $\left[
\begin{array}{c}
\bW \\
\bL
\end{array}
\right ]$ \corb{is assumed to be unitary}.

The first step is to filter out the effect of the trend by
\corb{projecting the observations} onto the orthogonal subspace.  Let
$\bZ_{\bW}: = \bW \bZ$, thus from equation
\eqref{Introduction:vectormodel} it follows that $\bZ_{\bW} = {\bf W}
({\bX \bbeta}+ {\boldsymbol \varepsilon}) = {\bf W{\boldsymbol
    \varepsilon}}$. Notice that the trend component ${\bX} \bbeta$ is
removed from the data ${\bf Z}$. The new log-likelihood function for
$\bZ_{\bW}$ becomes
\begin{equation}
  \begin{split}
\ell_{\bW}(\btheta)
&=-\frac{n}{2}\log(2\pi)-\frac{1}{2}\log
\det\{\bC_{\bW}(\btheta)\} 
-\frac{1}{2}\bZ_{\bW}\T\bC_{\bW}(\btheta)^{-1}\bZ_{\bW},
\end{split}
\label{Introduction:multilevelloglikelihood}
\end{equation}
where $\bC_{\bW}(\btheta) := \bW \bC(\btheta) \bW \T$ and
$\bZ_{\bW}\sim \mcN_{N-p}(\0,$ $\bC_{\bW}(\btheta))$.  A consequence
of the filtering is that we obtain an unbiased estimator
\cite{Castrillon2015}.  The decoupling of the likelihood function is
not the only advantage of using $\bC_{\bW}(\btheta)$. The following
theorem also shows that $\bC_{\bW}(\btheta)$ is more numerically
stable than $\bC(\btheta)$.

\begin{proposition} 
\label{Multilevelapproach:theo1}
Let $\kappa(A) \rightarrow \R$ be the condition number of the matrix
$A \in \R^{N \times N}$ then
$\kappa(\bC_{\bW}(\btheta)) \leq \kappa(\bC(\btheta))$.
\end{proposition}

Proposition \ref{Multilevelapproach:theo1} states that the condition
number of $\bC_{\bW}(\btheta)$ is less or equal to the condition
number of $\bC(\btheta)$. Thus computing the inverse of
$\bC_{\bW}(\btheta)$ (using a direct or iterative method) will
generally be more stable.
In practice, computing the inverse of $\bC_{\bW}(\btheta)$ can be
significantly more stable than $\bC(\btheta)$ depending on the choice
of $\mcQ^d_w$. This has many significant implications as it will now
be possible to solve numerically unstable problems.

There are other advantages to the structure of the matrix
$\bC_{\bW}(\btheta)$.  In section \ref{errorestimates} it is shown
that for a good choice of the $\mcP^{p}(\bbS)$ the entries of
$\bC_{\bW}(\btheta)$ decay rapidly, and most of the entries can be
safely eliminated. A level dependent criterion approach is shown in
Section \ref{MultilevelCovarianceMatrix} that indicates which entries
are computed and which ones are not. With this approach a sparse
covariance matrix $\tilde{\bC}_{\bW}$ can be constructed such that it
is close to $\bC_{\bW}$ in a matrix norm sense, even if the
observations are highly correlated with distance.

\subsection{Binary multilevel basis}
\label{MultilevelREML}

In this section the construction of Multilevel Basis (MB) is shown.
The approach followed in this section is a based on the MB
construction in \cite{Castrillon2013}. The MB can then be used to: (i)
form the multilevel likelihood
\eqref{Introduction:multilevelloglikelihood}; (ii) sparsify the
covariance matrix $\bC_{\bW}(\btheta)$; and (iii) improve the
numerical stability of the covariance matrix $\bC(\btheta)$ in it's
multilevel form. But first, let us establish notations and
definitions:
\begin{inparaenum}[i)]

\item For any index $i,j \in \mathbb{N}_{0}$, $1 \leq i \leq N$, $1
  \leq j \leq N$, let $\bve_{i}[j] = \delta[i-j]$, where
  $\delta[\cdot]$ is the discrete Kronecker delta function.
  
\item Let $\phi(\bx,\by;\btheta):\R^{d} \times \R^{d} \rightarrow \R$
  be the covariance function and assumed to be a positive definite.
  Let $\bC(\btheta)$ be the covariance matrix that is formed from all
  the interactions between the observation locations $\bbS$
  i.e. $\bC(\btheta) := \{ \phi(\bx_i,\by_j;\btheta) \}$, where $i,j,
  = 1,\dots,N$. \corb{We shall assume that the covariance function can
    be restricted to the following form: There exists a function
    $\varphi : [0,\infty) \rightarrow \R$ such that
      $\phi(\bx,\by;\btheta) = \phi(r;\btheta) := \varphi(r)$, where
      $r(\btheta) = ( (\bx-\by)^{T}$ $\text{diag}(\btheta) (\bx - \by)
      )^{\frac{1}{2}}$, $\btheta=[\theta_1, \dots, \theta_d] \in
      \R^{d}_{+}$, and $\text{diag}(\btheta) \in \R^{d \times d}$ is a
      diagonal matrix with the vector $\btheta$ on the diagonal.}
\end{inparaenum}

\begin{definition} \corb{Denote the Mat\'{e}rn covariance function:}
\begin{equation*}
\corb{\phi(r;\btheta)=\frac{1}{\Gamma(\nu)2^{\nu-1}} \left(
\sqrt{2\nu}\frac{r}{\rho} \right)^{\nu} K_{\nu} \left(
\sqrt{2\nu}\frac{r}{\rho} \right),}
\end{equation*}
where with a slight abuse of notation $\Gamma$ is the gamma function,
$r \in \R_{+}$, $0 < \nu$, $0 < \rho < \infty$, and $K_{\nu}$ is the
modified Bessel function of the second kind. It is understood from
context when $\Gamma$ is the gamma function.
\end{definition}

\begin{remark} The Mat\'{e}rn covariance function is a good choice for
the random field model. The parameter $\rho$ controls the length
correlation and the parameter $\nu$ changes the shape. For example, if
$\nu = 1/2 + n$, where $n \in \bbN_{+}$, then (see
\cite{abramowitz1964})
$\phi(r;\rho) = \exp   \bigg(-\frac{\sqrt{2\nu}r}{\rho} \bigg)
\frac{\Gamma(n + 1)}{\Gamma(2n + 1)}$
$\sum_{k = 1}^{n} \frac{(n+1)!}{k!(n-k)!}
\bigg(
\frac{ \sqrt{8v} r }{ \rho } 
\bigg)^{n-k}$
and $\nu \rightarrow \infty \Rightarrow \phi(r;\btheta) \rightarrow
\exp \bigg(-\frac{r^2}{2\rho^2} \bigg)$. Note that even for a moderate
number of derivatives the number of terms will grow exponentially fast
leading to a very complex expression. This motivates the study of
complex analytical extensions of the covariance function. See Section
\ref{errorestimates} for more details.
\label{multilevelapproach:remark1}
\end{remark}
\corb{The first step is to decompose the observation locations of
  $\bbS$ in the hypercube domain $\Gamma^{d}$ into a multilevel domain
  decomposition.} A good choice is based on the a kD-tree
decomposition of the space $\R^{d}$ \cite{Dasgupta2008}.  Other
choices include Projection (RP) trees. \corb{Kd-trees are usually
  applied for searching algorithms such as range and nearest
  neighbor. Kd-tree is a particular case of a binary tree, which are
  also used as decisions trees, sorting and classification, among many
  other applications.}

\corb{We refer the reader to \cite{deBerg2008} for the construction of
  the kd-tree. However, usually a kd-tree has at most one location
  point for each of the leaves. Instead, the leaf is set to a maximum
  of $p$ observations for the version that is used in this paper.}
First start with the root zero node and corresponding to cell
$B^{0}_{0}$ at level $0$ that contains all the observation nodes in
$\bbS$. Now, split these nodes into two children cells $B^{1}_{1}$ and
$B^{1}_{2}$ at level $1$ according to the following rule:
\corb{\begin{inparaenum}[i)]
\item Choose a unit vector $v$ in $\R^{d}$ along the axis of
  $\R^{d}$. This choice is the direction that leads to the maximum
  variance of the data in the cell along the direction of $v$.
\item Project all the nodes $\bx \in \bbS$ in the cell onto the unit
  vector $v$.
\item Split the cell with respect to the median
of the projections.
  \end{inparaenum}
  For each non empty cell $B^{k}_{l}$ with $\bbS_{T}$ points this
  procedure is repeated until the full binary tree is built.  This
  rule corresponds to Algorithm \ref{RPMLB:algorithm2-kd}. We now
  described the construction of the kd-tree by using this rule.}

\begin{algorithm}[h]
\footnotesize
\corb{
\begin{algorithmic}
  \Procedure{Rule}{$\bbS_{T}$}
  \State $vrs \gets $  variance of each coordinate direction of the data $\bbS_{T}$
  \State $v \gets $ direction of maximal entry of $vrs$
\State $proj\bbS \gets$ projection of $\bbS_T$ on $v$
  \State threshold $\gets $ median of $proj\bbS_T$
  \State \textbf{return} threshold, $v$, Rule($\bx$) $\gets \bx \cdot v  \leq$ threshold
  \EndProcedure
\end{algorithmic}
}
\caption{\corb{Rule function for cell-split}}
\label{RPMLB:algorithm2-kd}
\end{algorithm}

\corb{Algorithm \ref{RPMLB:algorithm1} initializes the tree by setting
  the node number and depth of the tree to zero. All of the original
  observations points $\bbS$ belong at node zero of the tree.  The
  MakeTree function from Algorithm \ref{RPMLB:algorithm2} is then
  executed to start the tree construction.}

\begin{algorithm}[h]
\footnotesize
\begin{algorithmic}
  \Procedure{InitialTree}{$\bbS$,$n_0$}
  \State node $\gets$ 0, depth $\gets$ 0
        \State Tree $\leftarrow$ MakeTree($\bbS$, node, depth + 1, $n_0$)
        \EndProcedure
\end{algorithmic}  
\caption{InitialTree  function}
\label{RPMLB:algorithm1}
\end{algorithm}

\corb{Algorithm \ref{RPMLB:algorithm2} splits the observation into
  binary cells at each level of the tree depth. Given the input
  observation locations in $\bbS_T$ they are split into two cells:
  Left and right according to the Rule. The tree is constructed by
  calling the MakeTree function recursively, both left and right. Note
  that the MakeTree function will construct all the left cells first
  until a leaf is reached. At this point the recursion is unwrapped
  one step and then the right cell is constructed. This is repeated
  many times over until all of the leafs are reached and the final tree
  is produced.}

\begin{algorithm}[h]
\footnotesize
\begin{algorithmic}

  \Procedure{MakeTree}{$\bbS_T$, node, depth, $n_0$}

         \State Tree.node $\gets$ node, Tree.depth $\gets$ depth - 1, node $\leftarrow$ node + 1
        \If {$ \lvert \bbS_T \rvert < n_0$} \textbf{return}
        \Comment{Leaf of the tree} 
        \EndIf
        \State (Rule, threshold, $v$) $\leftarrow$ ChooseRule($\bbS$)
        \State (Tree.LeftTree, node)  $\gets$ MakeTree($\bx \in \bbS_T$: Rule($\bx$) = True, node, depth + 1, $n_0$)
        \State (Tree.RightTree, node) $\gets$ MakeTree($\bx \in \bbS_T$: Rule($\bx$) = false,  node, depth + 1, $n_0$)
        \State Tree.threshold = threshold, Tree.$v$ = $v$
        \State \textbf{return} Tree

  \EndProcedure
  
\end{algorithmic}  
\caption{MakeTree  function}
\label{RPMLB:algorithm2}
\end{algorithm}

\corb{A binary tree is produced, which is of the form $B^{0}_{0}$,
  $B^{1}_{1}$, $B^{1}_{2}$, $B^{2}_{3}$, $B^{2}_{4}$, $B^{2}_{5}$,
  $B^{2}_{6}$, $\dots $, where $t$ is the maximal depth (level) of the
  tree. Note that each non zero cell $B^l_k$ will correspond to a
  particular node and depth number. In Figure \ref{MLRLE:fig1}
  an example illustration of the kd-tree is shown with a maximal
  set of locations at the leaves set to four.}

  \corb{Now, let $\mcB$ be the set of all the cells in the tree and
    $\mcB^{n}$ be the set of all the cells at level $0 \leq n \leq
    t$. In addition, for each cell a unique node number, current tree
    depth, threshold level and projection vector are also assigned to
    the node. In the Matlab code, this will be useful for searching
    the tree.  Algorithms \ref{RPMLB:algorithm2-kd},
    \ref{RPMLB:algorithm1}, and \ref{RPMLB:algorithm2} describe in
    more detail the construction of the kD-tree.}

\corb{Using the binary tree a multilevel basis for $\bbR^{N}$ is constructed.}
Suppose there is a one-to-one mapping between the set of unit
vectors $\mcE:=\{\bve_{1},\dots,$ $\bve_{N}\}$, which is denoted as
leaf unit vectors, and the set of locations $\{
\bx_{1},\dots,\bx_{N}\}$, i.e. $\bx_{n} \longleftrightarrow \bve_{n}$
for all $n = 1, \dots, N$. It is clear that the span of the vectors
$\{\bve_{1},\dots,\bve_{N}\}$ is $\bbR^{N}$.  \corb{The next step is to
construct a new basis of $\R^{N}$ that is multilevel and orthonormal.}

\setlength{\tabcolsep}{16pt}
\begin{figure*}
\begin{center}
  \begin{tabular}{c c} \begin{tikzpicture}\begin{scope} 
 [scale = 0.45,  every node/.style={scale=0.6}, place/.style={circle,draw=blue!50,fill=blue!20,thick,
     inner sep=0pt,minimum size=1.5mm}]

 \draw[step=8,gray,very thin] (0, 0) grid (8, 8);
    \draw (4,0) to (4,8);
    
    \draw (0,5) to (4,5);
    \draw (2.2,5) to (2.2,8);
    \draw (0,2) to (4,2);

    \draw (0,5) to (4,5);

    \draw (4,4.15) to (8,4.15);
    \draw (6.75,4.15) to (6.75,8);
    \draw (6,0) to (6,4.15);

    \node at (0.5,7.5) [place] {};
    \node at (0.3,6.3) [place] {};

    \node at (2.5,5.5) [place] {};
    \node at (3.2,5.2) [place] {};

    \node at (5,6) [place] {};
    \node at (3.8,5.5) [place] {}; \node at (3.8,6) [place] {};

    \node at (0.5,3.5) [place] {};
    \node at (1.5,2.5) [place] {};
    \node at (2.3,2.2) [place] {};
    
    \node at (1.3,0.3) [place] {};
    \node at (2.7,0.5) [place] {};
    \node at (2.2,1.4) [place] {};
    \node at (2.6,1.4) [place] {};
    \node at (4.2,3.5) [place] {}; \node at (3.7,3.3) [place] {};

    \node at (6.5,4.3) [place] {}; \node at (7.5,5) [place] {};

    \node at (4.3,2.3) [place] {};
    \node at (5.7,3.5) [place] {};
    \node at (6.2,3.4) [place] {};
    \node at (7.3,2.4) [place] {};

    \node at (7,7) [place] {};
    \node at (6,7.5) [place] {};
    \node at (7.5,7.5) [place] {};

    \node at (6.5,2.0) [place] {};
    \node at (0.5,7.0) [place] {};
    \node at (2.0,7.0) [place] {};
    \node at (5,3.75) [place] {}; \node at (6,7.0) [place] {};
    \node at (7,2.0) [place] {}; \node at (7.5,4.3) [place] {}; 

    \node at (7.5,8.5) [] {$B^{0}_0$};
    \node at (0.6,5.5) [] {$B^{3}_{7}$};
    \node at (3,7) [] {$B^{3}_{8}$};
  \end{scope}
\end{tikzpicture} 
&
\begin{tikzpicture}[scale=0.65]
\begin{scope}[xshift=5cm, yshift=4cm,
place/.style={circle,draw=blue!50,fill=blue!20,thick,
      inner sep=0pt,minimum size=1.5mm},
placer/.style={circle,draw=blue!50,
  preaction={fill=babyblue,fill opacity=0.5}, thick,inner
  sep=0pt,minimum size=1.5mm}, ]

\Tree [.\node[placer]{$B^{0}_{0}$}; 
             [.\node[placer]{$B^{1}_{1}$};
                    [.\node[placer]{$B^{2}_{3}$}; 
                           [.\node[placer]{$B^{3}_{7}$};]
                           [.\node[placer]{$B^{3}_{8}$};] 
                    ]       
                    [.\node[placer]{$B^{2}_{4}$}; 
                           [.\node[placer]{$B^{3}_{9}$};] 
                           [.\node[placer]{$B^{3}_{10}$};] 
                    ] 
             ]                                        
             [.\node[placer]{$B^{1}_{2}$};
                    [      [.\node[placer]{$B^{2}_{5}$};
                                  [.\node[placer]{$B^{3}_{11}$};] 
                                  [.\node[placer]{$B^{3}_{12}$};] 
                           ]
                           [.\node[placer]{$B^{2}_{6}$}; 
                                  [.\node[placer]{$B^{3}_{13}$};] 
                                  [.\node[placer]{$B^{3}_{14}$};] 
                           ] 
                                          ]]
] 

\end{scope}
\end{tikzpicture}
\end{tabular}
\end{center}
\caption{\corb{Multilevel domain decomposition of the observations locations
  example.  All of the initial points in $\bbS$ are split using
  Algorithms \ref{RPMLB:algorithm2-kd}, \ref{RPMLB:algorithm1}, and
  \ref{RPMLB:algorithm2} until at most 4 points are left in each
of the cells corresponding to the leaves of the binary tree.}}
\label{MLRLE:fig1}
\end{figure*}

\begin{enumerate}
\item Start at the maximum level of the random projection tree,
  i.e. $q = t$.
\item For each leaf cell $B^{q}_{k} \in \mcB^{q}$ assume without loss
  of generality that there are $s$ observations nodes $\bbS^{q}_{k}:=\{
  \bx_1, \dots, \bx_s \}$ with associated vectors $C_k^{q} := \{
  \bve_1, \dots, \bve_s \}$.
Denote $\mcC^{q}_{k}$ as the span of the vectors in $C_k^{q}$.
\begin{enumerate}

\item Let $\bphi^{q,k} _{j} := \sum_{\bve_i \in C^q_k} c^{q,k} _{i,j}
  \bve_i, \hspace{2mm} j=1, \dots, a;
\hspace{2mm} \bpsi^{q,k}_{j} := \sum_{\bve_i \in C^q_k} d^{q,k}_{i,j}
\bve_i$, $j=a+1, \dots, s$, where $c^{q,k}_{i,j}$,
$d^{q,k}_{i,j} \in \mathbb{R}$ and for some $a \in \mathbb{N}^{+}$. Note
that $a$ is unknown up to this point, but will be computed from the
data.  It is desired that the new discrete MB vector $\bpsi^{q,k}_{j}$
be orthogonal to $\mcP^{p}(\mathbb{S})$, i.e., for all $g \in \mcP^{
  p}(\mathbb{S})$:
\begin{equation}
\sum_{i=1}^{n} g[i] \bpsi^{q,k}_{j}[i] = 0
\label{hbconstruction:eqn1}
\end{equation}
\item Form the matrix $\mcM^{q,k} := \bX \T \bV^{q,k}$, where
  $\mcM^{q,k} \in \R^{p \times s}$, $\bV^{q,k} \in \R^{N \times s}$,
  and $\bV^{q,k}: = [\bve_1, \dots, \bve_i, \dots,\bve_s ]$ for all $\bve_i
  \in C_k^q$. Now, suppose that the matrix $\mcM^{q,k} $ has rank $a$
  and then perform the Singular Value Decomposition (SVD). Denote by
  $\bU \bD \bV $ the SVD of $\mcM^{q,k} $, where $\bU \in \R^{ p \times
    p}$, $\bD \in \R^{p \times s}$, and $\bV \in \R^{s \times s} $.

  \begin{remark} Note that in practice we only keep track of the
    non-zero elements of the vectors $\bve_1, \dots, \bve_s$. Thus the
    computational cost is reduced significantly. This is taken into
    account in the complexity analysis in Lemma
    \ref{MultilevelREML:lemma1} and \ref{MultilevelREML:lemma2}
  \end{remark}
  
\item Following the same argument as in \cite{Castrillon2015} but
  adapted to the kd-tree decomposition equation
  \eqref{hbconstruction:eqn1} is satisfied with the following choice
\begin{equation*}
  \left[ \begin{array}{ccc|ccc}
      c^{q,k}_{1,1} & \dots &c^{q,k}_{1,a} & d^{q,k}_{1,a+1} & \dots &d^{q,k}_{1,s} \\
      c^{q,k}_{2,1} & \dots &c^{q,k}_{2,a} & d^{q,k}_{2,a+1} & \dots &d^{q,k}_{2,s} \\
      \vdots & \vdots & \vdots & \vdots & \vdots & \vdots   \\
      c^{q,k}_{s,1} & \dots &c^{q,k}_{s,a} & d^{q,k}_{a+1,s} & \dots &d^{q,k}_{s,s}
    \end{array}
\right] := \bV.
\end{equation*}

\noindent For this choice the coefficient $a$ is equal to the number
of non-zero singular values. Thus the columns $a+1$, \dots, $s$ form
an orthonormal basis of the nullspace ${N_0}(\mcM^{q,k} )$. Similarly,
the columns $1,\dots, a$ form an orthonormal basis of $\R^s \backslash
{N_0}(\mcM^{q,k})$. Since the vectors in $C^q_k$ are orthonormal then
$\bphi^{q,k}_{1}, \dots, \bphi^{q,k}_a$, $\bpsi^{q,k}_{a+1}, \dots,$
$\bpsi^{q,k}_s$ form an orthonormal basis of $\mcC^{q}_{k}$.  Moreover
$\bpsi^{q,k}_{a+1}, \dots, \bpsi^{q,k}_s$ satisfy equation
\eqref{hbconstruction:eqn1}, i.e., are orthogonal to
$\mcP^{p}(\mathbb{S})$ and are locally adapted to the locations
contained in the cell $B^{q}_{k}$.

\item Denote by $D_k^{q,k}$ the collection of all the vectors
  $\bpsi^{q,k}_{a+1}, \dots, \bpsi^{q,k}_s$. Notice that the vectors
  $\bphi^{q,k}_{1}, \dots,$ $\bphi^{q,k}_a$, which are denoted with a
  slight abuse of notation as the scaling vectors, are {\it not}
  orthogonal to $\mcP^{p}(\mathbb{S})$. They need to be further
  processed.

\item Let $\mcD^{q}$ be the union of the vectors in $D^{q}_k$ for
  all the cells $B^{q}_k \in \mcB^{q}_{k}$. Denote by
  $W_{q}(\mathbb{S})$ as the span of all the vectors in $\mcD^{q}$.

\end{enumerate}

\item For any two sibling cells denote $B^{q}_{\tt{left}}$ and
  $B^{q}_{\tt{right}}$ at level $q$ denote $C^{q-1}_{\tilde k}$ as the
  collection of the scaling functions from both cells, for some index
  $\tilde k$.

\item Let $q: = q - 1$. If $B^{q}_{k} \in \mcB^{q}$ is a leaf cell
  then repeat steps (b) to (d). However, if $B^{q}_{k} \in \mcB^{q}$
    is not a leaf cell, then repeat steps (b) to (d), but replace the
    leaf unit vectors with the scaling vectors contained in $C^{q}_k$
    with $C^{q-1}_{\tilde k}$.

  \item When $q = -1$ is reached stop.

\end{enumerate}

\corb{When the algorithm stops a series of orthogonal subspaces
  $V_{0}(\bbS), W_{0}(\mathbb{S}),$ $\dots,W_{t}(\mathbb{S})$ (and
  their corresponding basis vectors) are obtained.} These subspaces
are orthogonal to $V_{0}(\mathbb{S}) : = span \{ \phi_{1}^{0}, \dots,
\phi_{p}^{0} \}$. Note that the orthonormal basis vectors of
$V_{0}(\mathbb{S})$ also span the space $\mcP^{p}(\mathbb{S})$.
\begin{remark}
Following Lemma 2 in \cite{Castrillon2013} it can be shown that
$\R^{N} = \mcP^{p}(\mathbb{S}) \oplus
W_{0}(\mathbb{S}) 
\oplus W_{1}(\mathbb{S})
\oplus \dots \oplus W_{t}(\mathbb{S}),$
Also, it can then be shown that at most $\mcO(Nt)$
computational steps are needed to construct the multilevel basis of
$\R^{N}$.
\end{remark}

From the basis vectors of the subspaces $\mcP^{p}(\mathbb{S})^{\perp}
= \cup_{i=0}^{t} W_{i}(\mathbb{S})$ an orthogonal projection matrix
$\bW:\R^{N} \rightarrow (\mcP^{p}(\mathbb{S}))^{\perp}$ can be built.
The dimensions of $\bW$ is $(N - p) \times N$ since the total number
of orthonormal vectors that span $\mcP^{p}(\mathbb{S})$ is
$p$. Conversely, the total number of orthonormal vectors that span
$\mcP^{p}(\mathbb{S})^{\perp}$ is $N-p$.  Let $\bL$ be a matrix where
each row is an orthonormal basis vector of $\mcP^{p}(\mathbb{S})$. For
$i = 0,\dots,t$ let $\bW_i$ be a matrix where each row is a basis
vector of the space $W_i(\mathbb{S})$. The matrix $\bW \in
\mathbb{R}^{(N - p) \times N}$ can now be formed, where $\bW := \left[
  \bW_t\T, \dots, \bW_0\T \right] \T$.  Following a similar approach
to Lemma 2.11 in \cite{Castrillon2013} it can be shown that:
i) The matrix $\bP := \left[
\begin{array}{c}
\bW \\
\bL
\end{array}
\right ]$ is orthonormal, i.e., $\bP\bP\T= \bI$.
ii) Any vector $\bv
\in \R^{n}$ can be written as $\bv = \bL\T\bv_{L} + \bW\T\bv_{\bW}$
where $\bv_{L} \in \R^{p} $ and $\bv_{\bW} \in \R^{N-p}$ are unique.
The following useful lemmas are proved:
\begin{lemma} Assuming that $n_0 < 2p$,
for any level $q=0,\dots,t$ there is at most $p2^{q}$ multilevel
basis vectors.
\label{MultilevelREML:lemma1}
\end{lemma}

\begin{lemma} Assuming that $n_0 < 2p$ for any level $q = 0, \dots, t$
  any multilevel vector $\bpsi^{q}_m$ associated with a cell $B^{q}_k
  \in \mcB^{q}$ has at most $2^{t-q+1} p$ non zero entries.
\label{MultilevelREML:lemma2}
\end{lemma}

From Lemma \ref{MultilevelREML:lemma1} and \ref{MultilevelREML:lemma2}
it can be shown that the matrix $\bW$ contains at most $\mcO(Nt)$
non-zero entries and $\bL$ contains at most $\mcO(Np)$ non-zero
entries. Thus for any vector $\bv \in \R^{n}$ the matrix vector
products $\bW \bv$ and $\bL \bv$ are respectively calculated with at
most $\mcO(Nt)$ and $\mcO(Np)$ computational steps.

\section{Multilevel covariance matrix}
\label{MultilevelCovarianceMatrix}

The multilevel covariance matrix $\bC_{\bW}(\btheta)$ and sparse
version $\tilde \bC_{\bW}(\btheta)$ can be now constructed.  Recall
from the discussion in Section \ref{multilevelapproach} that
$\bC_{\bW}(\btheta):=\bW \bC(\btheta) \bW \T$. From the multilevel
basis construct in Section \ref{MultilevelREML} the following operator
is built: $\bW := \left[ \bW_t\T, \dots, \bW_0\T \right] \T$. Thus the
covariance matrix $\bC(\btheta)$ is transformed into
$\bC_{\bW}(\btheta)$, where each of the blocks
$\bC^{i,j}_{\bW}(\btheta) = \bW_i \bC(\btheta) \bW_j \T$ are formed
from all the interactions of the MB vectors between levels $i$ and
$j$, for all $i,j = 0, \dots, t$. The structure of
$\bC_{\bW}(\btheta)$ is shown in Figure \ref{multilevelcov:fig2}(a).
Thus for any $\bpsi^{i}_{\tilde{l}}$ and $\bpsi^{j}_{\tilde{k}}$
vectors there is a unique entry of $\bC^{i,j}_{\bW}$ of the form
$(\bpsi^{i}_{\tilde{k}})\T \bC(\btheta) \bpsi^{j}_{\tilde{l}}$.  In
Section \ref{errorestimates} we show that far field entries of
$\bC_{\bW}(\btheta)$, i.e. $(\bpsi^{i}_{\tilde{k}})\T \bC(\btheta)
\bpsi^{j}_{\tilde{l}}$, decay sub-exponentially with respect to
$p(d,w)$ if there exists an analytic extension of the covariance
function on a well defined domain in $\bC^{d}$. Thus it is not
necessary to compute all the entries. We introduce a distance
criterion approach to produce a sparse matrix $\tilde
\bC_{\bW}(\btheta)$.

\subsection{Sparsification of multilevel covariance matrix}

A sparse version of the covariance matrix $\bC_{\bW}(\btheta)$ can be
built by using a level and distance dependent strategy:
\begin{inparaenum}[i)]
\item Given a cell $B^{i}_{k}$ at level $i \geq 0$ identify the
  corresponding tree node value Tree.node and the tree depth
  Tree.depth. Note that the Tree.depth and the MB level $q$ are the
  same for $q = 0,\dots,t$.
\item Let $\bbK \subset \bbS$ be all the observations nodes contained
  in the cell $B^{i}_{k}$.
\item Let $\tau_{i,j} \geq 0$ be the distance parameter given by the
  user corresponding to the level $i,j$ from the block
  $\bC^{i,j}_{\bW}(\btheta)$.
\item Let the Targetdepth be equal to the desired level of the tree.
\end{inparaenum}

The objective now is to find all the cells at the Targetdepth that
    overlap a hyper rectangle which is extended from $B^{i}_{k}$.  For
    all observations $\bx \in B^{i}_{l}$ along each dimension $k = 1,
    \dots, d$ let $x^{min}_k := \min_{ x_k \in B^i_m} x_k$ and
    $x^{max}_k := \max_{ x_k \in B^i_m} x_k$.  Any cell that
    intersects the interval $[x^{min}_{k} - \tau_{i,j} ,x^{max} +
      \tau_{i,j}]$ is included. This is done by searching the tree
    from the root node. At each traversed node check that all the
    nodes $\bx \in \bbK$ satisfy the following rule: If
$\bx \cdot \mbox{Tree}.v + \tau_{i,j} \leq$ Tree.threshold
then search down the left tree. If 
$\bx \cdot \mbox{Tree}.v - \tau_{i,j} >$ Tree.threshold.
the search down the right tree. Otherwise search both trees.
The full search algorithm is described in Algorithms
\ref{MLCM:algorithm3}, \ref{MLCM:algorithm4}, and \ref{MLCM:algorithm5}.

\begin{algorithm}[h]
  \footnotesize
\begin{algorithmic}
  \Procedure{SearchTree}{Tree, $\bbK$, Targetdepth, $\tau_{i,j}$}
  \State Targetnodes $\gets \emptyset $, Targetnodes $\gets$ LocalSearchTree(Tree, $\bbK$, Targetdepth, $\tau_{i,j}$, Targetnodes)
  \State \textbf{return} Targetnodes
  \EndProcedure  
\end{algorithmic}  
\caption{SearchTree function}
\label{MLCM:algorithm3}
\end{algorithm}

\begin{algorithm}[h]
  \footnotesize
\begin{algorithmic}
  \Procedure{LocalSearchTree}{Tree, $\bbK$, Targetdepth, $\tau_{i,j}$, Targetnodes}
  \If{Targetdepth = Tree.depth}
      \State \textbf{return}
      Targetnodes $\gets$ Targetnodes $\cup$ Tree.node
  \EndIf
  \If{Targetdepth = Leaf} \textbf{return}
  \EndIf
  \State LeftRule  =  ChooseLeftRule($\bbK$, Tree, $\tau_{i,j}$), RightRule =  ChooseRightRule($\bbK$, Tree, $\tau_{i,j}$)
\State Targetnodes $\leftarrow$ LocalSearchTree(Tree.LeftTree, 
   $\bbK$, Targetdepth, $\tau_{i,j}$, Targetnodes)
\State
   Targetnodes $\leftarrow$ LocalSearchTree(Tree.RightTree, 
   $\bbK$, Targetdepth, $\tau_{i,j}$, Targetnodes)
\State \textbf{return} Targetnodes
  \EndProcedure
\end{algorithmic}  
\caption{LocalSearchTree function}
\label{MLCM:algorithm4}
\end{algorithm}

\begin{algorithm}[h] 
\footnotesize
  \begin{algorithmic}
  \Procedure{ChooseLeftRule}{$\bbK$,Tree,$\tau_{i,j}$}
  \State  \textbf{return}  Rule($\bx$) := $\bx \cdot \mbox{Tree}.v + \tau_{i,j} \leq$ Tree.threshold
  \EndProcedure
    \Procedure{ChooseRightRule}{$\bbK$,Tree,$\tau_{i,j}$}
  \State  \textbf{return}  Rule($\bx$) := $\bx \cdot \mbox{Tree}.v + \tau_{i,j} >$ Tree.threshold
  \EndProcedure  
\end{algorithmic}  
\caption{ChooseLeftRule
  and
  ChooseRightRule rule functions
}
\label{MLCM:algorithm5}
\end{algorithm}

\begin{figure*}[ht]
\begin{center}

\raisebox{0cm}{
  \begin{tikzpicture} 
  \begin{scope}[thick,scale=0.25, every node/.style={scale=0.5}]
    [place/.style={circle,draw=blue!50,fill=blue!20,thick,
      inner sep=0pt,minimum size=1.5mm}]

    \filldraw[fill=babyblue,semitransparent, 
      thick] (0, 0) rectangle (16, 16);

    \draw[thin] (8,0) to (8,16);
    \draw[thin] (0,8) to (16,8);
    \draw[thin] (14,0) to (14,16);
    \draw[thin] (0,2) to (16,2);
    \draw[thin] (12,0) to (12,16);
    \draw[thin] (0,4) to (16,4);

    \node at (15,1) [] {${\bf G}_{\bW}$ };
    \node at (10,12) [] {$\bC^{t,t-1}_{\bW}(\btheta)$};
    \node at (4.5,6) [] {$\bC^{t-1,t}_{\bW}(\btheta)$};
    \node at (4.5,1) [] {$\bC^{0,t}_{\bW}(\btheta)$};
    \node at (10,6) [] {$\ddots$};
    \node at (13,1) [] {$\dots$};
    \node at (15,3) [] {$\vdots$};
    \node at (4.5,12) [] {$\bC^{t,t}_{\bW}(\btheta)$};

    \node at (8.3,-1.5) {\Large (a)};
\end{scope}
  \end{tikzpicture}
  }
 \begin{tikzpicture}[scale=0.5, every node/.style={scale=0.5}]
  \begin{scope} 
    [place/.style={circle,draw=blueish,fill=blueish,
        inner sep=0pt,minimum size=1.5mm},
      placegray/.style={circle,draw=gray!50,fill=gray!20,
        inner sep=0pt,minimum size=1.5mm},
        placenew/.style={circle,draw=darkorange!75,fill=darkorange!75,
      inner sep=0pt,minimum size=1.5mm}]
    \draw[step=8,gray,very thin] (0, 0) grid (8, 8);
    
    \node at (14.5,3.88) [] {\includegraphics[trim = 14.5cm 2cm 10cm 1.75cm,
        clip=true,
        height=8.43cm]{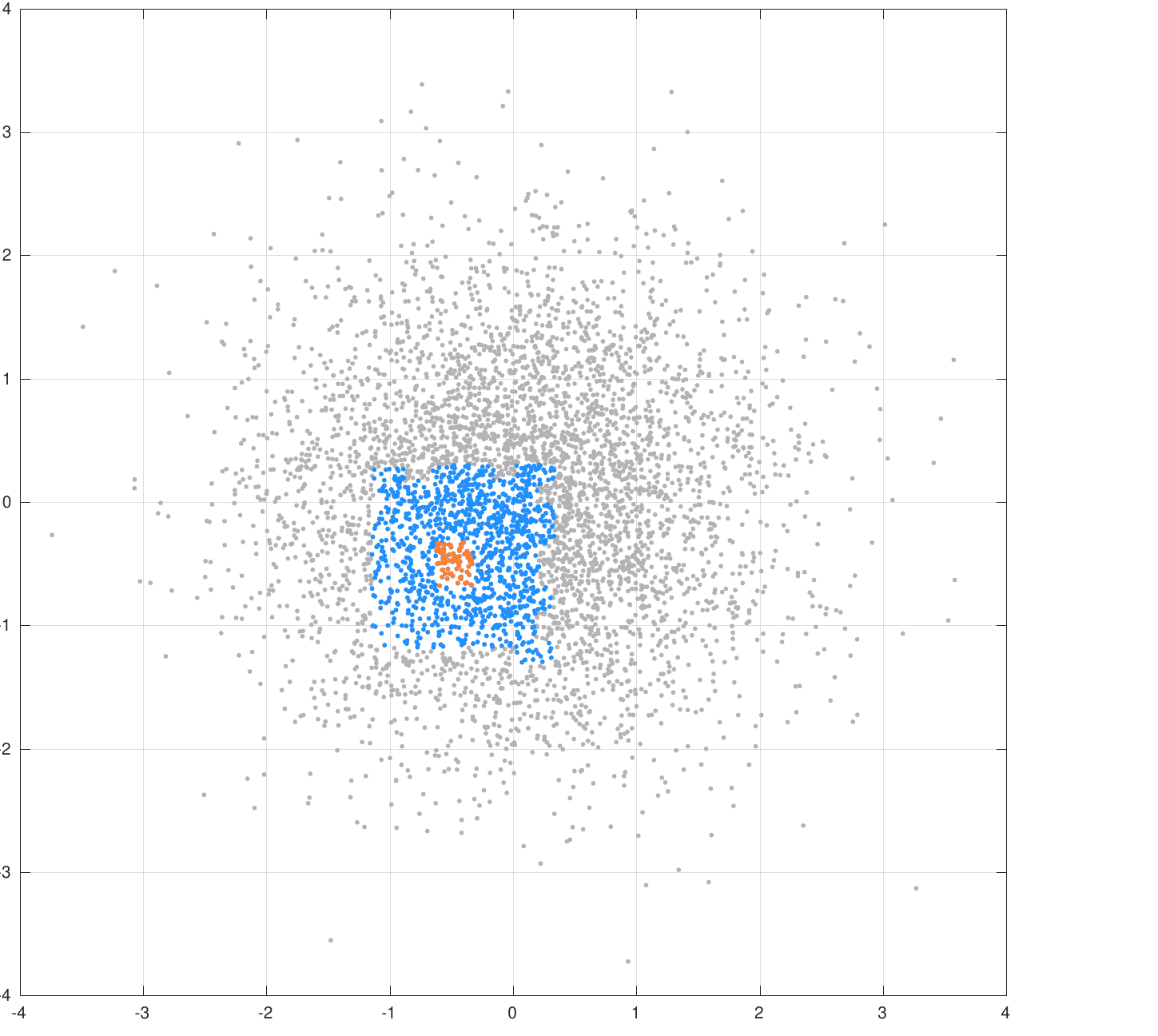}};
    
    \draw (2.2,5) to (2.2,8);
    \draw (0,5) to (2.2,5);
    \draw (6,4.15) to (8,4.15);
    \draw (6,0) to (6,4.15);

    \draw (4,0) to (4,8);
    \draw (0,5) to (4,5);
    
    \draw (0,2) to (4,2);
    \draw (0,5) to (4,5);
    \draw (4,4.15) to (8,4.15);
    \draw (6.75,4.15) to (6.75,8);
        
    \draw[dashed,gray] (6,4.15) to (6,8);
    \draw[dashed,gray] (0,2.3) to (6,2.3);
  
    \node at (0.5,7.5) [placenew] {};
    \node at (0.3,6.3) [placenew] {};
    
    \node at (2.5,5.5) [place] {};
    \node at (3.2,5.2) [place] {};

    \node at (5,6) [place] {};
    \node at (3.8,5.5) [place] {}; \node at (3.8,6) [place] {};   

    \node at (0.5,3.5) [place] {};
    \node at (1.5,2.5) [place] {};
    \node at (2.3,2.2) [place] {};
    
    \node at (1.3,0.3) [placegray] {};
    \node at (2.7,0.5) [placegray] {};
    \node at (2.2,1.4) [placegray] {};
    \node at (2.6,1.4) [placegray] {};
    \node at (4.2,3.5) [place] {}; \node at (3.7,3.3) [place] {};

    \node at (6.5,4.3) [place] {}; \node at (7.5,5) [placegray] {};

    \node at (4.3,2.3) [place] {};
    \node at (5.7,3.5) [place] {};
    \node at (6.2,3.4) [placegray] {};
    \node at (7.3,2.4) [placegray] {};

    \node at (7,7) [placegray] {};
    \node at (6,7.5) [place] {};
    \node at (7.5,7.5) [placegray] {};

    \node at (6.5,2.0) [placegray] {};
    \node at (0.5,7.0) [placenew] {};
    \node at (2.0,7.0) [placenew] {};
    \node at (5,3.75) [place] {}; \node at (6,7.0) [place] {};
    \node at (7,2.0) [placegray] {}; \node at (7.5,4.3) [placegray] {}; 

    \node at (4,8.75) [] {\Large $\tau_{i,j}$};
    \node at (-1,4.25) [] {\Large $\tau_{i,j}$};

    \draw[dashed,gray] (2,6.3) to (2,8);
    \draw[dashed,gray] (0,6.3) to (2,6.3);
    
    \coordinate (A) at (2,8);
    \coordinate (B) at (6,8);
    \coordinate (C) at (0,6.3);
    \coordinate (D) at (0,2.3);

\draw[dim={,10pt}]  (A) --  (B);
\draw[dim={,-15pt}]  (C) --  (D);

\node at (4,-0.65) {\Large (b)};
\node at (14.5,-0.65) {\Large (c)};
  \end{scope}
\end{tikzpicture}

\end{center}

\caption{Multilevel matrix and Neighborhood identification from source
  cell on a random kD-tree decomposition of observation locations in
  $\R^{2}$. (a) Multilevel covariance matrix where ${\bf G}_{\bW}
  :=\bC^{0,0}_{\bW}(\btheta)$.  (b) Cartoon example of axis wise
  distance criterion $\tau_{i,j}$ using Algorithms
  \ref{MLCM:algorithm3}, \ref{MLCM:algorithm4}, and
  \ref{MLCM:algorithm5}. The orange observations knots correspond to
  the source cell. The blue knots correspond to all the target
  nodes. The gray knots are not included in the list of target nodes.
  (c) Example of local neighborhood contained in the axis wise
  distance $\tau_{i,j}$.  The orange nodes are contained in the source
  cell. The blue nodes are are contained in the local neighborhood
  cells. The grey dots are all the observations that are not part of
  the source or local neighborhood cells.}
\label{multilevelcov:fig2}
\end{figure*}

In Figure \ref{multilevelcov:fig2} (b) \& (c) an example for searching local
neighborhood cells of randomly placed observations in $\R^{2}$ is
shown. The orange nodes correspond to the source cell. By choosing a
suitable value for $\tau_{i,j}$ the blue nodes in the immediate cell
neighborhood are found by using Algorithms \ref{MLCM:algorithm3},
\ref{MLCM:algorithm4}, and \ref{MLCM:algorithm5}.
The sparse matrix blocks $\bC^{i,j}_{\bW}(\btheta)$ can be built from
all the cells that are obtained from SearchTree function of Algorithm
\ref{MLCM:algorithm5}. Compute all the entries of
$\bC^{i,j}_{\bW}(\btheta)$ that correspond to the interactions between
any two cells $B^{i}_k \in \mcB^{i}$ and $B^{j}_l \in \mcB^{j}$. In
Algorithm \ref{MLCM:algorithm6}) the construction of the sparse matrix
$\tilde \bC^{i,j}_{\bW}(\btheta)$ is shown.

\begin{remark}
Since the matrix $\tilde \bC_{\bW}(\btheta)$ is symmetric it is only
necessary to compute the blocks $\bC^{i,j}_{\bW}(\btheta)$ for $i = 1,
\dots, t$ and $j = i, \dots t$.
\end{remark}

\begin{algorithm}[h]
  \footnotesize
\begin{algorithmic}
  \Procedure{Construction}{Tree, $i$, $j$, $\tau_{i,j}$, $\mcB^i$,
              $\mcB^j$, $\mcD^i$, $\mcD^i$, $\bC(\btheta)$}
  \State Targetnodes $\leftarrow \emptyset$
        \For{$B^{i}_{m} \in \mcB^{i}$}
             \State $\bbK \leftarrow B^{i}_{m}$
             \For{$B^{j}_{q} gets $
            SearchTree(Tree, $\bbK$, Targetdepth $(i)$, $\tau_{i,j}$, 
            Targetnodes)}
                \For{$\psi^i_k \in D^{i}$}
                     \For{$\psi^j_l \in D^{j}$}
                         \State Compute $(\bpsi^{i}_{k})\T \bC(\btheta) \bpsi^{j}_{l}$
                          in $\tilde \bC^{i,j}_{\bW}(\btheta)$
                     \EndFor
                \EndFor
             \EndFor
        \EndFor
  \EndProcedure
\end{algorithmic}  
\caption{Construction of sparse matrix $\tilde \bC^{i,j}_{\bW}(\btheta)$}
\label{MLCM:algorithm6}
\end{algorithm}

\subsection{Computational cost of the multilevel
  matrix blocks of $\tilde{\bC}_{\bW}$}

The cost of computing the multilevel blocks $\tilde{\bC}^{i,j}_{\bW}$
will in general be $\mcO(N^2)$. However, for the special case that $d
= 2$ and $d = 3$ it is possible to use a fast summation method such as
the Kernel Independent Fast Multipole Method (KIFMM) by
\cite{ying2004} to compute the blocks more efficiently. To my
knowledge, there exists no equivalent fast summation method in higher
dimensions that works satisfactorily.
This KIFMM algorithm is flexible and efficient for computing the
matrix vector products $\bC(\btheta)\bx$ for a large class of kernel
functions, including the Mat\'{e}rn covariance function.  Given
$\tilde N$ sources and $\tilde M$ targets, experimental results show a
computational cost of about $\mcO(\tilde N + \tilde M)$, $\alpha
\approx 1$ with good accuracy ($\varepsilon_{FMM}$ between $10^{-6}$
to $10^{-8}$) with a slight degrade in the accuracy with increased
source nodes.

\begin{asum} Let $\bA(\btheta) \in \R^{\tilde M \times \tilde N}$ be a kernel
matrix formed from $\tilde N$ source observation nodes and $\tilde M$
target nodes in the space $\R^{d}$.  Suppose that there exists a fast
summation method that computes the matrix-vector products
$\bA(\btheta)\bx$ with $\varepsilon_{FMM}>0$ accuracy in $\mcO((\tilde
N + \tilde M)^{\alpha})$ computations, for some $\alpha \geq 1$ and
any $\bx \in \R^{d}$.
\end{asum}

For the kD-tree it is not possible to determine a-priori the sparsity
of the blocks $\tilde{\bC}^{i,j}_{\bW}(\btheta)$.
However, for a given a value $\tau_{i,j} \geq 0$ by running Algorithm
\ref{MLCM:algorithm3} on every cell $B^{i}_k \in \mcB^{i}$, at level
$i$, with the Targetdepth corresponding for level $j$ it is possible
to determine the computational cost of constructing the sparse blocks
$\tilde{\bC}^{i,j}_{\bW}(\btheta)$ under the following assumption. Suppose
that maximum number of cells $B^{j}_k \in \mcB^{j}$ given by Algorithm
\ref{MLCM:algorithm3} is bounded by some $\gamma^{i,j} \in \bbN_+$.

\begin{proposition} 
  The cost of computing each block $\tilde{\bC}^{i,j}_{\bW}(\btheta)$
  for $i,j = 1,\dots,t$ by using a fast summation method with $1 \leq
  \alpha \leq 2$ is bounded by $\mcO(\gamma_{i,j} p 2^{i} ($
  $2^{t-j+1} p + 2^{t-i+1} p)^{\alpha} + 2p 2^{t})$.
\label{MultilevelREML:prop1}
\end{proposition}

\section{Multilevel estimator and predictor}
\label{multilevelestimator}
\corb{The multilevel decomposition, kd-tree and basis can be exploited
  in such a way to significantly reduce the computational burden and
  to further increase the numerical stability of the estimation and
  prediction steps.} This is an extension of the multilevel estimator
and predictor formulated in \cite{Castrillon2015} to binary trees in
higher dimensions. The former is based on Oct-tree decompositions,
thus making it unsuitable for higher dimensional problems.

\subsection{Estimator}
\corb{The multilevel likelihood function, $l_{\bW}(\theta)$ (see
  equation \eqref {Introduction:multilevelloglikelihood}), has the
  clear advantage of being decoupled from the vector $\bbeta$.}
Furthermore, the multilevel covariance matrix $\bC_{\bW}(\btheta)$
will be more numerically stable than $\bC(\btheta)$ thus making it
easier to invert and to compute the determinant.  However, it is not
necessary to perform the MLE estimation on the full covariance matrix
$\bC_{\bW}(\btheta)$, instead construct a series of multilevel
likelihood functions $\tilde{\ell}^n_{\bW}(\btheta)$, \corb{for $n =
  0,\dots t-1$, by applying the partial transform $[\bW_t \T, \dots,
    \bW_{n} \T ]\T$ to the data $\bZ$.  The following likelihood
  functions are obtained: For $n = 0,\dots,t-1$}
\begin{equation}
  \begin{split}
\tilde{\ell}^n_{\bW}(\btheta)
=-\frac{\tilde{N}}{2}\log(2\pi)-\frac{1}{2}\log
\det\{\tilde{\bC}_{\bW}^n(\btheta)\}
-\frac{1}{2}(\bZ^n_{\bW})\T\tilde{\bC}^n_{\bW}(\btheta)^{-1}\bZ^n_{\bW},
\end{split}
\label{Introduction:multilevelloglikelihoodreduced2}
\end{equation}
where $\bZ^n_{\bW} :=[\bW_t \T, \dots, \bW_{n} \T ] \T \bZ$,
$\tilde{N}$ is the length of $\bZ^n_{\bW}$,
$\tilde{\bC}^n_{\bW}(\btheta)$ is the $\tilde{N} \times \tilde{N}$
upper-left sub-matrix of $\tilde{\bC}_{\bW}(\btheta)$ and
$\bC^n_{\bW}(\btheta)$ is the $\tilde{N} \times \tilde{N}$ upper-left
sub-matrix of $\bC_{\bW}(\btheta)$. \corb{For the case that $n = t$
  then
\begin{equation}
  \begin{split}
\tilde{\ell}^t_{\bW}(\btheta)
=-\frac{\tilde{N}}{2}\log(2\pi)-\frac{1}{2}\log
\det\{\tilde{\bC}_{\bW}^t(\btheta)\}
-\frac{1}{2}(\bZ^t_{\bW})\T\tilde{\bC}^t_{\bW}(\btheta)^{-1}\bZ^t_{\bW},
  \end{split}
  \label{Introduction:multilevelloglikelihoodreduced3}
\end{equation}
where $\bZ^t_{\bW} := \bW_t  \bZ$, $\tilde{N}$
is the length of $\bZ^t_{\bW}$, $\tilde{\bC}^t_{\bW}(\btheta)$ is the
$\tilde{N} \times \tilde{N}$ upper-left sub-matrix of
$\tilde{\bC}_{\bW}(\btheta)$ and $\bC^t_{\bW}(\btheta)$ is the
$\tilde{N} \times \tilde{N}$ upper-left sub-matrix of
$\bC_{\bW}(\btheta)$.}

\corb{A consequence of this approach is that
  for $n = 0, \dots, t$ the matrices $\bC^n_{\bW}(\btheta)$ are
  increasingly more stable, thus easier to solve computationally, as
  shown in the following theorem.}
\begin{proposition} 
  \label{Multilevelapproach:theo2}
Let $\kappa(A) \rightarrow \R$ be the condition number of the matrix
$A \in \R^{N \times N}$ then
$\kappa(\bC^{t}_{\bW}(\btheta)) \leq \kappa(\bC^{t-1}_{\bW}(\btheta)) \leq 
\dots
\leq
\kappa(\bC_{\bW}(\btheta)) \leq
\kappa(\bC(\btheta)).$
\end{proposition}

\begin{remark} If $\bC(\btheta)$ is symmetric positive definite 
then for $n = 0, \dots, t$ the matrices $\bC^{n}_{\bW}(\btheta)$ are
symmetric positive definite. The proof is immediate.
\corb{Furthermore, for $n = 1,\dots,t$, if the matrix $\tilde
  \bC^{n}_{\bW}(\btheta)$ is close to $\bC^{n}_{\bW}(\btheta)$, in
  some matrix norm sense, then the condition number of $\tilde
  \bC^{n}_{\bW}(\btheta)$ will be close to $\bC^{n}_{\bW}(\btheta)$.}
Full error bounds will be derived in a future publication.
\end{remark}

\subsection{Predictor}

\corb{In this section, we demonstrate how to construct a multilevel BLUP
with a well-conditioned multilevel covariance matrix. Furthermore, the
multilevel predictor is exact, implying that the solutions of the
multilevel predictor and the BLUP equations \eqref{GLSbeta} and
\eqref{KrigBLUP} are identical. The key insight is to recognize that
the BLUP arises from a constrained optimization problem (see Section
\ref{Introduction}). By seeking the solution within the constrained
space, it becomes possible to formulate a set of equations that are
numerically more stable. The multilevel approach bypasses the need to
invert the covariance matrix $\bC(\btheta)$, thereby
avoiding the challenges posed by ill-conditioned matrices. It's
important to note that ill-conditioned matrices offer no guarantees of
numerical accuracy, as previously discussed in \cite{Golub96}.}

Consider the following system of equations
\begin{eqnarray}
\left( {{\begin{array}{*{20}c}
 \bC(\btheta) \hfill & \bX \hfill \\
 \bX\T \hfill & \0 \hfill \\
\end{array} }} \right)\left( {{\begin{array}{*{20}c}
 \hat \bgamma \hfill \\
 \hat \bbeta \hfill \\
\end{array} }} \right)=\left( {{\begin{array}{*{20}c}
 \bZ \hfill \\
 \0 \hfill \\
\end{array} }} \right).
\label{Spatial:problem}
\end{eqnarray}
From the argument given in \cite{Nielsen2002} it is not hard to show
that the solution of this problem leads to equation \eqref{GLSbeta}
and $\hat \bgamma(\btheta) = \bC^{-1}(\btheta)(\bZ - \bX \hat
\bbeta(\btheta))$. The BLUP can be evaluated as
\begin{equation}
  \hat Z(\bx_0)
  =\bk(\bx_0)\T\hat \bbeta(\btheta)+\bc(\btheta)\T
  \hat \bgamma(\btheta)
\label{Spatial}
\end{equation}
and the Mean Squared Error (MSE) at the target point $\bx_0$ is given by
$1 + 
\tilde{\bu}
\T
(\bX \T 
\bC(\btheta)^{-1} \bX )^{-1}
\tilde{\bu}$
$-\bc(\btheta)\T\bC^{-1}(\btheta)\bc(\btheta)$
where $\tilde{\bu}\T := (\bX \bC^{-1}(\btheta)
\bc(\btheta) - \bk(\bx_0))$.

From \eqref{Spatial:problem} it is observed that $\bX\T \hat
\bgamma(\btheta) = \0$. \corb{This implies that $\hat{\bgamma} \in
  \R^{n} \backslash \mcP^{p}(\mathbb{S})$ i.e. the solution for
  $\hat{\bgamma}$ lives in a lower dimensional space, and can be
  written as $\hat{\bgamma} = \bW\T \bgamma_{\bW}$ for some
  $\bgamma_{\bW} \in \R^{N-p}$. From equation \eqref{Spatial:problem},
  rewrite $\bC(\btheta) \hat \bgamma + \bX \hat \bbeta = \bZ$ as}
\begin{equation}
\bC(\btheta) \bW\T \bgamma_{\bW} + \bX \hat \bbeta =
       \bZ.
\label{Spatial:eqn1}
\end{equation}
Now apply the matrix $\bW$ to equation \eqref{Spatial:eqn1} and obtain
$\bW \{\bC(\btheta) \bW\T \bgamma_{\bW} + \bX \hat \bbeta\} = \bW
\bZ$. \corb{Since the columns of $\bX$ belong to
  $\mcP^{p}(\mathbb{S})$ then $\bW \bX = \0$ and therefore}

\corb{
\begin{equation}
    \bC_{\bW}(\btheta) \bgamma_{\bW} =
      \bZ_{\bW}.
\label{Spatial:eqn2}
  \end{equation}
  Solving these set of equations leads to the unique solution $\hat
  \bgamma_W$ and the vector $\hat \bgamma$ can be obtained by applying
  the inverse transform $\bW\T$ i.e.  $\hat \bgamma = \bW\T
  \bgamma_{\bW}$.  From \eqref{Spatial:problem} the GLS $\hat \bbeta$
  can now be computed as
\begin{equation}
\hat{\bbeta} = (\bX\T \bX)^{-1}\bX \T (\bZ -
\bC(\btheta)\hat{\bgamma}).
\label{Spatial:eqn3}
\end{equation}
Thus $\hat \bgamma(\btheta)$ are $\bbeta(\btheta))$ obtained by
solving the multilevel equations \eqref{Spatial:eqn2} and
\eqref{Spatial:eqn3}, which also solves the system of equations
\eqref{Spatial:problem}. Thus the BLUP is solved exactly.}

\corb{
  \begin{remark}
    Solving for $\hat Z(\bx_0)$, $\hat \gamma (\btheta)$,
    $\bbeta(\btheta)$ only requires the indirect inversion of the
    covariance matrix $\bC_{\bW}$ in equation
    \eqref{Spatial:eqn3}. This is in contrast to the classical BLUP
    method (See Remark \ref{Introduction:remark1}).
\label{Spatial:remark1}
    \end{remark}
  \begin{remark}
  Notice that to solve the GLS estimate $\hat \bbeta$ it is not
  necessary to compute the full GLS of equation \eqref{GLSbeta}, but a
  least squares is all that is required. This is in contrast to the
  GLS estimate of equation \eqref{GLSbeta} where if an iterative
  method is used the covariance matrix $\bC(\btheta)$ has to be
  inverted for each of the columns of $\bX$ i.e. $p$ times.
  \end{remark}
  \begin{remark}
A simple preconditioner $\bP_{\bW}$ can be formed from the diagonal
entries of the matrix $\bC_{\bW}$ i.e. $\bP_{\bW} = diag(\bC_{\bW})$ leading
to the following system of equations
$\bP_{\bW}^{-1}\bC_{\bW}(\btheta)
\bgamma_{\bW} = \bP_{\bW}^{-1} \bZ_{\bW}$.
Note that in some cases $\bC_{\bW}(\btheta)$ will have very small
condition numbers. For this case we can set $\bP_{\bW}:= I$, i.e. no
preconditioner.
    \end{remark}
\begin{theorem}
If the covariance function $\phi:\Gamma_d \times \Gamma_d \rightarrow
\R$ is positive definite, then the matrix $\bP_{\bW}(\btheta)$ is always
symmetric positive definite.
\label{multilevelSpatial:lemma2}
\end{theorem}
}

\section{Numerical computation of multilevel estimator and predictor}
\label{numericalcomputation}

\subsection{Estimator: Computation of 
$\log{\det\{ \tilde{\bC}^n_{\bW}\}}$ and $(\bZ^n_{\bW})\T
  (\tilde{\bC}^n_{\bW})^{-1}\bZ^n_{\bW}$}

An approach to computing the determinant of
$\tilde{\bC}^n_{\bW}(\btheta)$ is to apply a sparse Cholesky
factorization technique such that $\bG\bG\T =
\tilde{\bC}^n_{\bW}(\btheta)$, where $\bG$ is a lower triangular
matrix. Notice that the eigenvalues of $\bG$ are located on the
diagonal. This leads to $\log \det \{\tilde{\bC}^n_{\bW}(\btheta)\} =
2 \sum_{i = 1}^{\tilde{N}} \log{\bG_{ii}}$.

The direct application of the sparse Cholesky algorithm can lead to
significant fill-in of the factorization matrix $\bG$. To alleviate
this problem it is typical to use matrix reordering techniques. In
particular, the fill-in are reduced by using the sparse Cholesky
factorization \emph{chol} from the Suite Sparse 4.2.1 package
(\cite{Chen2008,Davis2009,Davis2005,Davis2001,Davis1999}) coupled with
Nested Dissection (NESDIS) function package.  In practice, this
approach leads to a significant reduction of fill-in. A theoretical
worse case complexity bounded exists for $d = 2,3$ dimensions (see
\cite{Castrillon2015}).

\corb{There are two choices for the computation of $(\bZ^n_{\bW}) \T
  \tilde \bC^n_{\bW}(\btheta)^{-1}$:} i) a Cholesky factorization of
$\tilde{\bC}^n_{\bW}(\btheta)$, or ii) a Preconditioned Conjugate
Gradient (PCG).  The PCG choice requires significantly less memory and
allows more control of the error.  However, the sparse Cholesky
factorization of $\tilde{\bC}^n_{\bW}(\btheta)$ has already been used
to compute the determinant. Thus we can use the same factors to
compute $ (\tilde \bZ_{\bW}^n) \T\tilde{\bC}^n_{\bW}(\btheta)^{-1}
\tilde \bZ^n_{\bW}$.  The PCG avenue will be explored further in
Section \ref{comppred}.

\subsection{Predictor computation}
\label{comppred}

For the predictor stage a different approach is used. Instead of
inverting the sparse matrix $\tilde \bC_{\bW}(\btheta)$ a
Preconditioned Conjugate Gradient (PCG) method is employed to compute
$\hat \bgamma_{\bW} = \bC_{\bW}(\btheta)^{-1} \bZ_{\bW}$.

Recall that $\bC_{\bW} = \bW \bC(\btheta) \bW \T$, $\hat \bgamma_{\bW} =
\bW \hat \bgamma$ and $\bZ_{\bW} = \bW \bZ$. Thus the matrix vector
products $\bC_{\bW}(\btheta) \bgamma_{\bW}^n$ in the PCG iteration are computed
within three steps:
$\bgamma_{\bW}^n \xrightarrow[(1)]{\bW \T \bgamma_{\bW}^n} 
\ba_n \xrightarrow[(2)]{\bC(\btheta) \ba_n}
\bb_n \xrightarrow[(3)]{\bW \bb_n}
\bC_{\bW}(\btheta) \bgamma_{\bW}^n $,
where $\bgamma_{\bW}^0$ is the initial guess and $\bgamma_{\bW}^n$ is the $n^{th}$
iteration of the PCG.
\begin{inparaenum}[(1)]
\item Transformation from multilevel representation to single
  level. This is done in at most $\mcO(Nt)$ steps.
\item Perform matrix vector product using a summation method. For $d =
  2,3$ a KIFMM is used to compute the matrix vector products with
  $\alpha \approx 1$. For $d > 3$ to my knowledge there is no reliable
  fast summation method.
\item Convert back to multilevel representation.
\end{inparaenum}

The matrix-vector products $\bC_{\bW}(\btheta) \bgamma_{\bW}^n$, where
$\bgamma_{\bW}^n \in \R^{N-p}$, are computed in $\mcO(N^{\alpha} + 2
Nt)$ computational steps to a fixed accuracy $\varepsilon_{FMM} > 0$.
Note that $\alpha \geq 1$ is dependent on the efficiency of the fast
summation method. The total computational cost is $\mcO(kN^{\alpha} +
2Nt)$, where $k$ is the number of iterations needed to solve
$\bP^{-1}_{\bW} \bC_{\bW}(\btheta) \bar{\bgamma}_{\bW} (\btheta) =
\bP^{-1}_{\bW} \bar{\bZ}_{\bW}$ to a predetermined accuracy
$\varepsilon_{PCG} > 0$.

\begin{remark}
The introduction of a preconditioner can degrade the accuracy for
computing $\hat \bgamma_{\bW} = \bC_{\bW}(\btheta)^{-1} \bZ_{\bW}$
with the PCG method. \corb{The residual accuracy $\varepsilon_{PCG}$ of the
PCG iteration has to be set such that the residual of
the \emph{unpreconditioned} system $\|\bC_{\bW}(\btheta) \bgamma_{\bW}
(\btheta) - \bZ_{\bW}\|_{l^2} < \varepsilon$ for a user given
tolerance $\varepsilon > 0$.}
\end{remark}

Now compute $\hat \bgamma = \bW\T \hat \bgamma_{\bW}$ and $\hat{\bbeta} =
(\bX\T \bX)^{-1}\bX \T (\bZ - \bC(\btheta)\hat{\bgamma})$ in at most
$\mcO(N^{\alpha} + Np + p^{3})$ computational steps. The matrix vector
product $\bc(\btheta)\T \hat{\bgamma}(\btheta)$ is computed in
$\mcO(N)$ steps.  Finally, the total cost for computing the estimate
$\hat{\bZ}(\bx_0)$ from \eqref{Spatial} is $\mcO(p^{3} + (k +
1)N^{\alpha} + 2Nt)$.

\section{Multilevel covariance matrix decay}
\label{errorestimates}

We derive decay estimates of the multilevel covariance
matrix. \corb{This section is somewhat technical and can be skipped on
  a first read of the paper.}  It can be shown that most of the
coefficients are small and thus it is not necessary to compute all of
them. The final objective is to build a posteriori error estimates for
$\bx_{\bW} = \bC^{n}_{\bW}(\btheta)^{-1}\bZ^n_{\bW}$ and \corb{$\log
  \det( \bC^{n}_{\bW}(\btheta))$} that are needed for solving the
multilevel estimator MLE.  However, the full analysis is extensive and
will be completed in a future publication. As a first step we show the
decay of the multilevel covariance matrix. Note that this is not
trivial and uses the results derived in the supplement. We recommend
to first read the appendix since part of the notation used in this
section is defined there. \corb{However, some of the notation and
  definitions will be included in this section so as to make it more
  self contained.}

\corb{In the paper \cite{Castrillon2015} the authors derive the decay
  rates of the entries of the covariance matrix for multilevel
  matrices in $\R^{d}$ based on Taylor's theorem. This approach is
  well suited for a small number of dimensions $d$. However, as $d$
  increases the number of derivatives in the Taylor's theorem
  increases combinatorially. Furthermore, the dimension of the domain
  of these derivatives increases with respect to $d$. For large
  dimensional problems it becomes increasingly difficult to compute
  the constants that depend on the derivatives in the bounds derived
  in \cite{Castrillon2015}. In contrast, by using the complex analytic
  approach the constants can be uniformly bounded and depend on the
  region of the analytic extension.  This is the reason that in the
  field of uncertainty quantification for stochastic Partial
  Differential Equations with high dimensional random parameters
  complex analyticity is used instead
  \cite{nobile2008a,nobile2008b,Castrillon2016,Castrillon2021a,Castrillon2021b}.
  We follow this approach.}

\corb{The decay of the coefficients of the matrix $\bC_{\bW}(\btheta)$
  will depend directly on the choice of the multivariate index set
  $\mcQ^{d}_{w}$ and the complex analytic regularity extension of the
  covariance function.  In general, the Mat\'{e}rn covariance function
  will be analytic except for a derivative discontinuity at the
  origin. However, with the application of the distance criterion
  $\tau_{i,j} > 0$ a minimal distance can be guaranteed and the origin
  can be avoided all together.  In the following theorem, without loss
  of generality, it is assumed that the covariance function $\bphi$ is
  defined on the domain $\Gamma^{d} \times \Gamma^{d}$ for on any two
  cells $B^{i}_{m} \in \mcB^{i}$ and $B^{j}_{q} \in \mcB^{j}$. This is
  achieved by using a pullback that we shall explain shortly.
  Furthermore, we restrict our attention to any two cells $B^{i}_{m}$
  and $B^{j}_{q}$ that do not overlap. This will guarantee that the
  center of the covariance function is avoided and the existence of
  complex analytic extension as shown in Theorem
  \ref{errorestimates:theorem2}.}

\corb{Suppose that $\sigma > 0$ and denote by
\begin{equation*}
\begin{split}
  \mcE_{\sigma} &:= \Big\{
  z \in \bbC, \sigma \geq
\delta \geq 0 :\,\Real{z} = \frac{e^{\delta} + e^{-\delta}
}{2}cos(\theta), 
\Imag{z} = \frac{e^{\delta} 
  - e^{-\delta}}{2}sin(\theta), \\
&\theta \in [0,2\pi)
  \Big\}
\end{split}
  \end{equation*}
as the region bounded by a Bernstein ellipse. Thus $\mcE_{\sigma}$ is
an extension into the complex plane from the domain $\Gamma \equiv
[-1,1]$ (see Figure \ref{erroranalysis:sparsegrid:polyellipse}).  Let
$\mcE_{\sigma,n} \subset \bbC^{d}$ a complex region bounded by a
Bernstein ellipse such that the restriction on $\Gamma_{d}$ is along
the $n^{th}$ dimension and form the polyellipse $\mcE^{d}_{\sigma}:=
\prod_{n=1}^{d} \mcE_{\sigma,n}$.}

\begin{theorem} Suppose that $0< \delta < 1$, $\hat
  \sigma := \sigma (1 - \delta)$, and
  \corb{$\phi(\balpha,\bgamma;\btheta) \in C^{0}(\Gamma^{d} \times
    \Gamma^{d};\R)$, where $\balpha,\bgamma \in \Gamma^{d}$,}
  can be analytically extended on
  $\mcE^{d}_{\sigma} \times \mcE^{d}_{\sigma}$ and is bounded by
  $\tilde M(\phi)$. Let $\mcP^{p}(\mathbb{S})^{\perp}$ be the subspace
  in $\R^{N}$ generated by the index set $\mcQ^{d}_{w}$ for some $w
  \in \bbN_{+}$. For $i,j = 0,\dots,t$ consider any multilevel vector
  $\bpsi^{i}_m \in \mcP^{p}(\mathbb{S})^{\perp}$, with $n_m$ non-zero
  entries, from the cell $B^{i}_{m} \in \mcB^{i}$ and any multilevel
  vector $\bpsi^{j}_{q} \in \mcP^{ p}(\mathbb{S})^{\perp}$, with $n_q$
  non-zero entries, from the cell $B^{j}_{q} \in \mcB^{j}$. \corb{If
    $B^{i}_{m}$ and $B^{j}_{q}$ do not overlap, and $p(d,w) \geq
    \left(\frac{2 d}{\kappa(d)}\right)^{d}$ then \corb{$\lvert \sum_{k
        = 1}^{N} \sum_{h = 1}^{N} \phi(\bx_k,\by_h; \btheta)
      \bpsi^i_m[h] \bpsi^j_q[k] \rvert$} is less or equal to}
\begin{equation*}
\begin{split}
\sqrt{n_mn_q}
\left( \frac{
  C(\tilde M,\sigma)^{d} e^{d - \sigma(1 - \delta) + 1} \hat \sigma d
}
 {
   (\sigma \delta)^{d}} \right)^2
 \left( \frac{e^{\hat \sigma}}{1 - e^{-\hat \sigma}} \right)^{2d}
 \exp \left(-\frac{2d}{e} \hat \sigma  p^{\frac{1}{d}}
 \right) p^{2\left(\frac{d-1}{d}\right)}.
\end{split}
\end{equation*}
\label{errorestimates:theorem1}
\end{theorem}

\begin{figure}[t]
\centering
\includegraphics[trim = 0 130 0 130,
  clip,scale = 0.4]{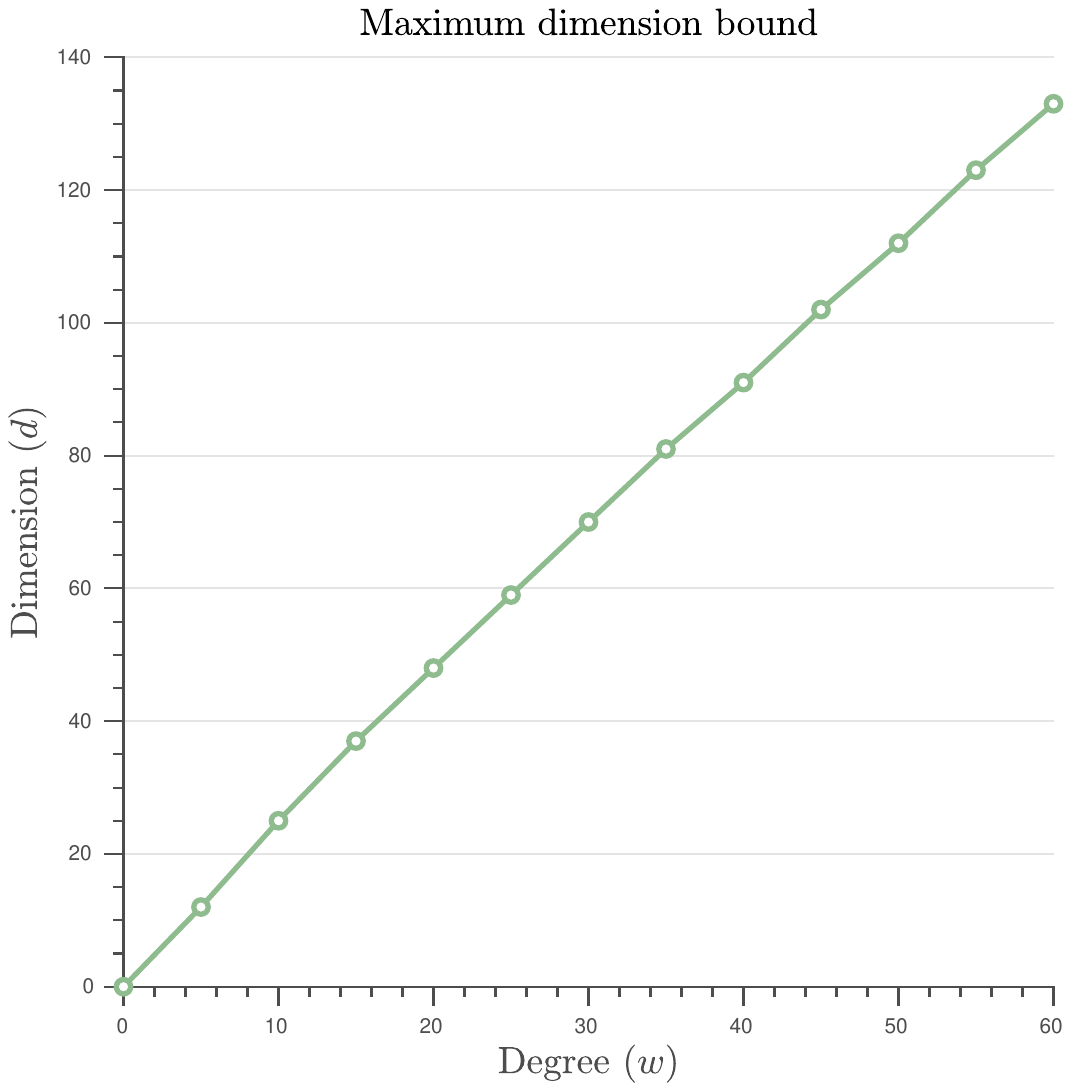}
\caption{\corb{Validity of Theorem \ref{errorestimates:theorem1}
    maximum dimensionality with respect to the degree $w$ given by the
    bound $p(d,w) \geq \left(\frac{2 d}{\kappa(d)}\right)^{d}$.}}
\label{errorestimates:Fig1}
\end{figure}

\corb{In Figure \ref{errorestimates:Fig1} a plot of the validity of
  the bound given by Theorem \ref{errorestimates:theorem1} is
  shown. For example, for $w = 20$ Theorem
  \ref{errorestimates:theorem1} will be valid for problems of up to $d
  = 60$ dimensions. For a small number of dimensions $d$ the bounds
  derived in \cite{Castrillon2015} are sufficient. However, as the
  number of dimensions increases it is preferable to use complex
  analyticity and the bound from Theorem \ref{errorestimates:theorem1}.}

\begin{remark} Recall that the restriction $p(d,w) \geq \left(\frac{2
  d}{\kappa(d)}\right)^{d}$ is not strict and can be relaxed such that
  sub-exponential convergence is still obtained. See Remark
  \ref{interpolation:remark1}.
\end{remark}

\begin{remark}
  \corb{Note that the bounded given by Theorem \ref{errorestimates:theorem1}
  does not give an explicit bound with respect to the distance
  criterion $\tau_{i,j}$.  However, the size of the polyellipse
  $\mcE^{d}$ and magnitude of $\tilde M(\phi)$ on $\mcE^{d}$ will
  depend on $\tau_{i,J}$.  In Theorem \ref{errorestimates:theorem2}
  this dependence is shown explicitly.}
  \end{remark}

\begin{remark}
  The decay of the coefficients of $\bC^{i,j}_{\bW}$ is
  sub-exponential with respect to $p$.  Even for a moderate magnitude
  for $\hat \sigma > 0$, $p > 0$ and $d \geq 1$ the entries of the
  multilevel matrix $\bC^{i,j}_{\bW}$ that do not correspond to the
  cells given by the distance criterion parameter $\tau_{i,j} \geq 0$
  will be close to zero.
\end{remark}

Theorem \ref{errorestimates:theorem1} provides a mechanism to control
the decay of the coefficients of the multilevel covariance matrix
$\bC_{\bW}$. \corb{To apply Theorem \ref{errorestimates:theorem1} we
  need to show that there exists a complex analytic extension of the
  Mat\'{e}rn covariance function
\begin{equation*}
\phi(r;\btheta):=\frac{1}{\Gamma(\nu)2^{\nu-1}} \left(
\sqrt{2\nu} r(\btheta) \right)^{\nu} K_{\nu} \left(
\sqrt{2\nu} r(\btheta) \right)
\end{equation*}
and a uniform bound $\tilde M(\phi) \leq \infty$ on a subdomain in
$\bbC^{d} \times \bbC^{d}$, where $r(\btheta) = ( (\bx-\by)^{T}$
$\text{diag}(\btheta) (\bx - \by) )^{\frac{1}{2}}$,
$\btheta=[\theta_1, \dots, \theta_d] \in \R^{d}_{+}$ are positive
constants, and $\text{diag}(\btheta) \in \R^{d \times d}$ is a
diagonal matrix with the vector $\btheta$ on the diagonal for all
$\bx, \by \in \R^{d}$.}

\corb{Due to the low regularity at the center of the covariance function
$\phi(r;\btheta)$ a complex analytic extension will not
exist. However, if we avoid the center then such extension is
possible. Consider any multilevel vector $\bpsi^{i}_m \in
\mcP^{p}(\mathbb{S})^{\perp}$, with $n_m$ non-zero entries, from the
cell $B^{i}_{m} \in \mcB^{i}$ and any multilevel vector $\bpsi^{j}_{q}
\in \mcP^{ p}(\mathbb{S})^{\perp}$, with $n_q$ non-zero entries, from
the cell $B^{j}_{q} \in \mcB^{j}$.  By placing the restriction that
$\tau_{i,j} > 0$ and the cells $B^i_{m}$ and $B^j_{q}$ do not
intersect then the covariance function on each of the locations in
$B^{i}_{m}$ will not cross the cell in $B^{j}_{q}$. Thus the low
regularity center is avoided.}

\corb{Our first step is to construct a pullback of the covariance
  function defined on the region covered by the cells $B^{i}_{m}$ and
  $B^{j}_{q}$ onto the region $\Gamma^{d}\equiv [-1,1]^{d}$. The
  rational behind this is that the interpolation theory in Appendix
  \ref{PolynomialAppendix} and in \cite{Griebel2016} is defined on the
  domain $\Gamma^{d}$, thus we need to re-scale the covariance function
  on each of the cells $B^{i}_{m}$ and $B^{j}_{q}$.}

For $k = 1, \dots, d$ let $x^{min}_k := \min_{ x^*_k \in B^i_m}
x^*_k$, $x^{max}_k := \max_{ x^*_k \in B^i_m} x^*_k$, $y^{min}_k :=
\min_{ y^*_k \in B^i_m} y^*_k$, $y^{max}_k$ $:= \max_{ y^*_k \in
  B^i_m}$ $y^*_k$ and $\alpha_k,\gamma_k \in [-1,1]$. Define the
region $\mcX^{i}_{m} := [x^{min}_1,$ $x^{max}_1] \times \dots \times
[x^{min}_d,x^{max}_d]$ and $\mcY^{j}_{q} := [y^{min}_1,y^{max}_1]
\times \dots \times [y^{min}_d,y^{max}_d]$.

The next step is to redefine $\phi(\bx,\by;\btheta):\mcX^{i}_m \times
\mcY^{i}_{q} \rightarrow \R$ as
$\phi(\balpha,\bgamma;\btheta):\Gamma^d \times \Gamma^d \rightarrow
\R$ through a pullback. For $k = 1, \dots, d$, let $x_k =
\left(\frac{\alpha_k + 1}{2} \right) a_k + b_k$ and $y_k =
\left(\frac{\gamma_k + 1}{2} \right) c_k + d_k$, where $a_k =
x^{max}_{k} - x^{min}_{k}$, $b_k = x^{min}_{k}$, $c_k = y^{max}_{k} -
y^{min}_{k}$ and $d_k = y^{min}_{k}$. Thus, all of the locations
$\bbS$ in $B^{i}_{m}$ and $B^{j}_{q}$ will be contained in the
hyperectangles $\mcX^{i}_{m}$ and $\mcY^{j}_{q}$ respectively
\corb{Furthermore, we have also obtained the pullback from
$\mcX^{i}_{m}$ and $\mcY^{j}_{q}$ onto $\Gamma^{d}$}.

\corb{We now set certain parameters that insures the existence of the
  analytic extension from the dimensions of the hyperectangles
  $\mcX^{i}_m$ and $\mcY^{j}_q$. The Mat\'{e}rn function consists of
  polynomial and Bessel components.  The polynomial is an entire
  function, thus we do not have to worry about it.  However, the
  function $K_{\nu}(\vartheta)$ and $\vartheta^{\frac{1}{2}}$ are
  analytic for all $\vartheta \in \bbC$ except at the branch cut
  $(-\infty,0]$.  Thus it is sufficient to check the analytic
extension of $r(\btheta) = \Big( \sum_{k=1}^{d} \theta_{k} (x_k -
y_k)^{2} \Big)^{\frac{1}{2}}$.  The choice of these parameters will
allow us to construct a complex analytic extension that avoids the
branch cut at $(-\infty,0]$.  Please read the proof of Theorem
\ref{errorestimates:theorem2} for more details.}  Now, for all $k =
  1,\dots,d$ pick $\delta_{k} > 0 $ be such that $\delta_{k} \leq
\frac{\sqrt{32\,\tau_{i,j} ^2+8\,\tau_{i,j} +1}-1 - 4\,\tau_{i,j}
}{4\,\tau_{i,j}}$. Let $\bdelta := [\delta_1, \dots, \delta_k]$
and
\[
\xi(\btheta,\bdelta,\tau_{i,j}) := \tan^{-1}
\left(
\frac{
  \sum_{k=1}^{d} 2 \tau_{i,j}
\delta_{k} + 4 \tau^2_{i,j} \delta_{k}^2
}
{
\sum_{k=1}^{d} \theta_k 
  \left(
  \tau_{i,j}^2 (1 - 4 \delta_{k}^2) - \frac{\tau_{i,j}
    \delta_{k}}{2} \right)
}
\right).
\]

\corb{For $k = 1,\dots,d$ let $\sigma^{\alpha}_k := \cosh^{-1} \left(1
  + \frac{\tau_{i,j} \delta_{k}}{a_k} \right)$ and $\sigma^{\gamma}_k
  := \cosh^{-1} \left(1 + \frac{\tau_{i,j} \delta_{k}}{c_k}
  \right)$. From these parameters we can form the polyellipses
  $\mcE^{d}_{\alpha} := \prod_{k=1}^{d} \mcE_{\sigma^{\alpha}_k}$ and
  $\mcE^{d}_{\gamma} := \prod_{k=1}^{d} \mcE_{\sigma^{\gamma}_k}$ Note
  that each of the Bernstein ellipses will contain the closed interval
  $[-1,1]$, thus they are an extension of $\Gamma$ into the complex
  plane. From the choice of these parameters it is shown that there
  exists an analytic extension $r(\btheta)$ onto $\mcE^{d}_{\alpha}
  \times \mcE^{d}_{\gamma}$. The following result gives the existence
  of a complex extension of the Mat\'{e}rn kernel onto the
  polyellipses $\mcE^{d}_{\alpha} \times \mcE^{d}_{\gamma}$.}
\begin{theorem}
\corb{For any two cells $B^{i}_{m}$ and $B^{j}_{q}$ that do not
  overlap} with the associated distance criterion parameter
\corb{$\tau_{i,j} > 0$} let $\phi(\balpha,\bgamma;$ $\btheta):\Gamma^d
\times \Gamma^d \rightarrow \R$ be the pullback of the Mat\'{e}rn
covariance function $\phi(\bx,\by;\btheta):\mcX^{i}_m \times
\mcY^{j}_{q} \rightarrow \R$. Then there exists an analytic extension
of $\phi(\balpha,\bgamma;\btheta):\Gamma^d \times \Gamma^d \rightarrow
\R$ onto the polyellipse $ \mcE^{d}_{\alpha} \times \mcE^{d}_{\gamma}$
and
$  \lvert \phi(\cdot,\cdot;\btheta) \rvert \leq
\frac{  \left( 2 \nu \sum_{k=1}^{d} \theta_k \mcR(\delta_k,\tau_{i,j})
\right)^{\frac{\nu}{2}}
\lvert K_{\nu}
(\Xi(\btheta,\bdelta,\tau_{i,j}
) )\rvert}{
\Xi(\btheta,\bdelta,\tau_{i,j})^{\nu}
}$
\corb{on $\mcE^{d}_{\alpha} \times \mcE^{d}_{\gamma}$, where
\[
\Xi(\btheta,\bdelta,\tau_{i,j})
  :=
\lvert K_{\nu}\Big(\sqrt{\frac{\nu}{2}}
  \cos(\xi(\btheta,\bdelta,\tau_{i,j})/2)
\sum_{k=1}^{d} \theta_k 
  \Big(
  \tau_{i,j}^2 (1 - 4 \delta_{k}^2) - \frac{\tau_{i,j}
    \delta_{k}}{2} \Big)
  \Big) \rvert
 \]
  and
  $\mcR(\delta_k,\tau_{i,j}) :=
1 + \frac{9}{2}\tau_{i,j} \delta_{k} + 5 \tau^2_{i,j} \delta_{k}^2$.}
\label{errorestimates:theorem2}
\end{theorem}
\corb{We can now apply Theorem \ref{errorestimates:theorem2} to Theorem
\ref{errorestimates:theorem1} to obtain a bound on the coefficients of
$\bC_{\bW}(\btheta)$}.

\section{Numerical results}
\label{numericalresults}

The performance of the multilevel solver for estimation and prediction
formed from random datasets is tested. The results show that the
computational burden is significantly reduced while retaining good
accuracy. In particular, it is possible to now solve ill-conditioned
problems efficiently. The experimental setup is described in Section
\ref{experimental}.

\subsection{Condition numbers and sparsity of the covariance multilevel
  matrix}

\begin{figure*}[htpb]
\begin{center}
\begin{tikzpicture}[thick,scale=0.52, every node/.style={scale=0.52}]
  \node[inner sep=0pt] at (0,2.3)
  {

  \includegraphics[trim = 120 400 120 255,
      clip,width=4.4in,height=2in]{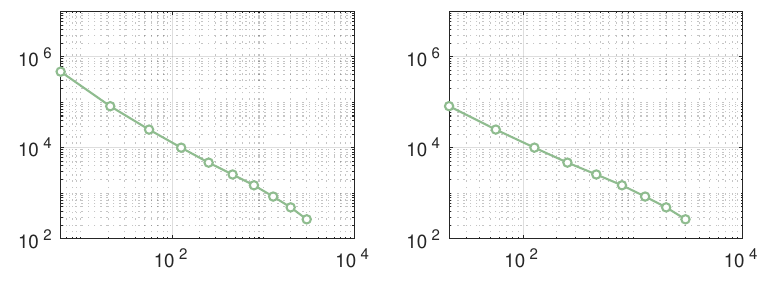}
    
  };
  \node[rotate = 90] at (-5.5,2.7) {$\kappa(\bC_{\bW}(\btheta))$};
  \node[rotate = 90] at (0,2.7) {$\kappa(\bC_{\bW}(\btheta))$};

    \node[inner sep=0pt] at (11.5,2.55)
  {

  \includegraphics[trim = 120 255 120 400,
      clip,width=4.4in,height=2in]{ConditionGraphsReduced.pdf}
    
  };  
  \node[rotate = 90] at (5.8,2.7) {$\kappa(\bC_{\bW}(\btheta))$};
  \node[rotate = 90] at (11.5,2.7) {$\kappa(\bC_{\bW}(\btheta))$};

\node at (-2.6,5.3) {
\begin{tabular}{c}
\small $Cube, d = 5, \bC^5_4, \rho=1$ \\
\small $\kappa(\bC(\btheta))= 1.1 \times 10^{7}$
\end{tabular}
};

\node at (2.9,5.3) {
\begin{tabular}{c}
\small $Cube, d = 5, \bC^5_4, \rho=10$ \\
\small $\kappa(\bC(\btheta))= 2.2 \times 10^{9}$
\end{tabular}
};

\node at (9,5.3)
{
\begin{tabular}{c}
\small $Sphere, d = 5,  \bC^5_4, \rho=1$ \\
\small $\kappa(\bC(\btheta))= 2.6 \times 10^{7}$
\end{tabular}
};

\node at (14.5,5.3)
{
\begin{tabular}{c}
\small $Sphere, d = 5, \bS^5_4, \rho=10$ \\
\small $\kappa(\bC(\btheta))=7.8 \times 10^{9}$
\end{tabular}
};
\node at (2.9,0)
      {$p$
        };
\node at (-2.6,0)
      {$p$
      };

      \node at (9,0)
      {$p$
        };
\node at (14.5,0)
      {$p$
      };

\end{tikzpicture}
\end{center}
\caption{Condition number of the multilevel covariance matrix
  $\bC_{\bW}(\btheta)$ with respect to the size $p$ of the Total
  Degree (TD) polynomial space. The number of observations corresponds
  to 16,000 nodes generated on a hypercube or n-sphere of dimension $d
  = 5$. The covariance function is chosen to be Mat\'{e}rn with $\nu =
  1$ and $\rho=1,10$.  The condition number of the covariance matrix
  $\bC(\btheta)$ is placed on the top of each subplot for
  comparison. The MB is constructed from a kD-tree.  As expected, as
  $p$ increases with $w$ the condition number of $\bC_{\bW}(\btheta)$
  decreases significantly. This is consistent with Theorem
  \ref{Multilevelapproach:theo1}.}
\label{numericalresults:fig1}
\end{figure*}

For many practical cases the covariance matrix $\bC(\btheta)$ becomes
increasingly ill-conditioned for the Mat\'{e}rn covariance function as
$\rho$, $\nu$ and the number of observations are increased. This leads
to instability of the numerical solver. It is now shown how effective
Theorem \ref{Multilevelapproach:theo1} becomes in practice.  In
Figure \ref{numericalresults:fig1} the condition number of the
multilevel covariance matrix $\bC_{\bW}(\btheta)$ is plotted with
respect to the cardinality $p(w,d)$ of $\mcQ^d_w$ for different $w$
levels. The multilevel covariance matrix $\bC_{\bW}(\btheta)$ is built
from the random cube $\bC^{d}_{4}$ or n-sphere $\bS^{d}_{4}$
observations. The covariance function is set to Mat\'{e}rn with $\nu =
1$ and $\rho = 1,10$.  As the plots confirm the covariance matrix
condition number significantly improves with increasing level
$w$. This is in contrast with the large condition numbers of the
original covariance matrix $\bC(\btheta)$.  This is consistent with
Theorem \ref{Multilevelapproach:theo1}.

In Table \ref{numericalresults:table1} sparsity and construction wall
clock times of the sparse matrices $\tilde{\bC}^{i}_{\bW}(\btheta)$,
$i = t, t-1, \dots$, for various values of $i$ are shown.  The
polynomial space of the index set $\mcQ^d_w$ is restricted to TD on a
n-Sphere with $d = 10$ dimensions. The domain decomposition is formed
with a kD-tree. The level of the index set is set to $w = 7$, which
corresponds $p = 1001$. The covariance function is Mat\'{e}rn with
$\nu = 3/4$, $\rho = 3/4$. The distance criterion for each $(i,j)$
multilevel covariance matrix block is set to
$\tau_{i,j} := 2^{(t - i)/2}2^{(t - j)/2} \tau$,
for $i = 1, \dots, t$ and $j = 1 \dots, t$, where $\tau = 3 \times
10^{-6}$.

The first observation to notice is that all the sparse matrices
$\tilde{\bC}^{i}_{\bW}(\btheta)$, $i = t, t-1, \dots$ {\it are very
  well conditioned, thus numerically stable}. This is in contrast to
the original covariance matrices that are in general poorly
conditioned. The sparsity of $\tilde{\bC}^{i}_{\bW}(\btheta)$ and the
Cholesky factor $\bG$ are shown in columns 7 and 9. The construction
time $t_{con}$ of the $\tilde{\bC}^{i}_{\bW}(\btheta)$ is shown in
column 9. In column 5 $t_{ML}$ is the time required to build the
multilevel basis.  We observe that for large matrices the sparse
matrix $\tilde{\bC}^{i}_{\bW}(\btheta)$ are built efficiently.  It is
noted that the sparse matrices in Table \ref{numericalresults:table1}
are built with a direct summation method due to the dimensionality.

\setlength{\tabcolsep}{7pt}
\begin{table*}[htpb]
  \caption{Sparsity test on the matrices $\tilde{\bC}^{i}_{\bW}$, $i =
    t, t-1, \dots$.  The polynomial space of the index set $\mcQ^d_w$
    is restricted to TD on a n-Sphere with $d = 10$ dimensions. The
    domain decomposition is formed from a kD-tree. The level of the
    index set is $w = 7$, which corresponds $p = 1001$. The kernel
    function is Mat\'{e}rn with $\nu = 3/4$, $\rho = 3/4$ and $\tau :=
    3 \times 10^{-6}$. The first column is the number of random
    n-Sphere nodes. The second is the maximum level of the kD tree and
    $i$ is the level of the sparse matrix $\tilde{\bC}^{i}_{\bW}$. The
    fourth column is the condition number of $\tilde{\bC}^{i}_{\bW}$,
    which is excellent.  The fifth column is the size of the matrix
    $\tilde{\bC}^{i}_{\bW}$.  The seventh column, $t_{ML}$, is the
    total time for the construction of the multilevel basis. The
    eighth column is the sparsity of $\tilde{\bC}^{i}_{\bW}$.  The
    ninth column, $t_{con}$ is the total time for the construction of
    the matrix $\tilde{\bC}^{i}_{\bW}$. The tenth column is the
    sparsity of the Cholesky factor $\bG$ (with nested dissection
    reordering) of the sparse matrix $\tilde{\bC}^{i}_{\bW}$. The last
    column is the total time to compute the Cholesky factor $\bG$.}
  \footnotesize
\begin{center}
\begin{tabular}{ r r r r c c r r c r c r}
\multicolumn{1}{c}{$N$} &
\multicolumn{1}{c}{$t$} &
\multicolumn{1}{c}{$i$} &
\multicolumn{1}{c}{$\kappa(\tilde \bC_{\bW}^{i})$} &
\multicolumn{1}{c}{Size} &
\multicolumn{1}{c}{$t_{ML}$} &
\multicolumn{1}{c}{$nz$} &
\multicolumn{1}{c}{$t_{con}$} &
\multicolumn{1}{c}{$nz(\bG)$} &
\multicolumn{1}{c}{$t_{sol}$}
 \\ 
 \hline
32,000 & 4 & 4 &   5  & 15,984 &  46 &  6.3\% &   11 &  3.1\% &  1 \\
32,000 & 4 & 3 &   8  & 23,992 &  46 & 10.4\% &   30 &  5.2\% &  3 \\
32,000 & 4 & 2 &  13  & 27,996 &  46 & 15.6\% &   82 &  7.8\% &  7 \\
32,000 & 4 & 1 &  19  & 29,998 &  46 & 20.1\% &  190 & 10.4\% & 16 \\
32,000 & 4 & 0 &  23  & 30,999 &  46 & 25.7\% &  310 & 13.0\% & 17 \\
\hline
64,000 & 5 & 5 &   6  & 31,968 & 104 &  3.5\% &   21 & 1.8\%  &  3 \\
64,000 & 5 & 4 &  11  & 47,984 & 105 &  6.3\% &   90 & 3.1\%  & 12 \\
64,000 & 5 & 3 &  18  & 55,992 & 106 &  9.6\% &  270 & 5.0\%  & 18 \\
64,000 & 5 & 2 & 121  & 59,996 & 121 & 13.4\% & 624 & 6.7\%  & 34 \\
\hline
128,000 & 6 & 6 &  8  &  63,936 &  237 & 4.0 \% & 120  & 2.1 \% & 15 \\
128,000 & 6 & 5 & 17  &  95,968 &  237 & 5.5 \% & 378 & 6.7 \% & 140 \\
\end{tabular}
\end{center}
\label{numericalresults:table1}
\end{table*}

\subsection{Prediction numerical accuracy under ill-conditioning}
In this section the numerical accuracy of the ML BLUP solver is tested
for a series of highly ill-conditioned matrices and compared with the
traditional formulae. Due to the ill-conditioning of the covariance
matrices we will see in general that the traditional formulae cannot
solve the BLUP with accuracy. Recall the relationship between accuracy
and ill-conditioning \cite{Golub96}. This problem is circumvented by
using the multilevel approach.  For the covariance function the
Gaussian kernel $\phi(r,\theta) = \exp(-\frac{r^2}{2\theta^2})$ is
used, where $\theta$ controls the width of the kernel.  This kernel is
notorious for leading to covariance matrices with large condition
numbers.  The observation locations $\{\bx_1,\dots,\bx_N\}$ are
randomly sampled on a unit disk (2D) with $N = 200$.  The multilevel
basis is constructed with degree $w = 8$, i.e. $p = 45$. Let $\bx_k :=
[x^k_1, x^k_2, \dots x^k_N]$, then observations for $n = 1, \dots N$
are formed as $Z_n = 1 + \sin( \alpha x^n_1)\sin( \alpha x^n_2)$,
where $\alpha \in \R$. The BLUP target nodes are be computed on $N_T =
200$ randomly sampled locations $\{\bg_1,\dots \bg_{N_T} \}$ on the
unit disk. The BLUP is computed for each of the target points as $\hat
Z(\bg_k)=\bk(\bg_k)\T\hat \bbeta+\bc(\btheta)\T
\bC(\btheta)^{-1}(\bZ-\bX\hat \bbeta)$, where
$\bc(\btheta)=\cov\{\bZ,Z(\bg_k)\}\in \R^{n}$ and $k = 1,\dots N_T$.

Suppose the BLUP is computed using the traditional formulae i.e.
$\hat \bbeta(\btheta) = (\bX\T \bC(\btheta)^{-1} $ $\bX)^{-1}$
$\bX\T\bC(\btheta)^{-1}\bZ$ and $\hat \bgamma(\btheta) =
\bC^{-1}(\btheta)(\bZ - \bX \hat \bbeta(\btheta))$. Let $\hat
\bbeta_{D}$ be the GLS coefficients and $\hat \bZ_D \in \R^{N_T}$ the
BLUP at the target nodes using a traditional double precision
computer. Conversely $\hat \bbeta_{52}$ and $\hat \bZ_{52}$ are
computed using the symbolic toolbox of MATLAB with 52 digits
accuracy. Note that the symbolic toolbox direct inversion methods are
very slow and cannot be used for large matrices.  Now, the BLUP is
computed using the multilevel method. Let $\hat{\bbeta}_{\bW}$ be the
GLS coefficients and $\hat \bZ^{\bW}_{D}$ be the BLUP at the target
nodes with a double precision computer. The relative errors are
computed with respect to the 52 digit solution as \corb{$\bepsilon_{\hat
  \bbeta_{\bW}} := \frac{\|\hat \bbeta_{\bW} - \hat \bbeta_{52}\|}{\| \hat
\bbeta_{52}\|}$}, $\bepsilon_{\hat \bZ^{\bW}_{D}} := \| \hat
\bZ^{\bW}_{D} - \hat \bZ_{52}\| \| \hat \bZ_{52} \|^{-1}$,
$\bepsilon_{\hat \bbeta_{D}} := \|\hat \bbeta_{D} - \hat \bbeta_{52}\|
\| \hat \bbeta_{52}\|^{-1}$, and $\bepsilon_{\hat \bZ_{D}} := \| \hat
\bZ_{D} - \hat \bZ_{52}\| \| \hat \bZ_{52} \|^{-1}$.

In Table \ref{numericalresults:table5} the accuracy results for the
traditional formulae and the multilevel BLUP are tabulated. Observe
that the original covariance matrices are very ill-conditioned and is
reflected with high errors in the computation of the BLUP. This is in
contrast to the multilevel approach, which leads to much higher
accuracies and lower condition numbers of the covariance matrices.
Notice that some round off errors still affect the overall accuracy.
However, the multilevel method is significantly more robust that the
traditional BLUP formulae.

\begin{table*}[htbp]
\footnotesize
\begin{center}
\setlength{\tabcolsep}{4pt}
\begin{tabular} { r c c c c  c c c r r l l l}
  \multicolumn{1}{r}{$\alpha$} & $\theta$ & $\kappa (\bC)$ & $\kappa (\bC_{\bW})$ &
 $\epsilon_{\hat \bbeta_{\bW}}$
  & 
  $\epsilon_{\hat \bZ^{\bw}_{D}}$
  & 
  $\epsilon_{\hat \bbeta_{D}}$
  & 
  $\epsilon_{\hat \bZ_{D}}$
  \\
  \hline 
1    & 10 & $1.62 \times 10^{19}$ & $9.55 \times 10^{3}$  &  $1.15 \times 10^{-4}$ &  $1.96 \times 10^{-6}$   & $1.38 \times 10^2$ & $3.00 \times 10^2$  \\
  0.5  & 10 & $1.62 \times 10^{19}$ & $9.55 \times 10^{3}$  &  $1.67 \times 10^{-9}$ &  $1.91 \times 10^{-9}$   & $2.71 \times 10^0$ & $4.39 \times 10^0$  \\
  2    & 10 & $1.62 \times 10^{19}$ & $9.55 \times 10^{2}$  &  $2.77 \times 10^{-2}$  & $1.60\times 10^{-3}$   & $1.41 \times 10^{3}$ & $1.66 \times 10^{4}$ \\
  2    & 3 &  $7.92 \times 10^{19}$ & $9.18 \times 10^{2}$  &  $2.35\times 10^{0}$  & $5.79\times 10^{-4}$ & $1.37\times 10^{3}$ & $1.47\times 10^{0}$ \\
  0.5  & 3 &  $7.92 \times 10^{19}$ & $9.18\times 10^{2}$ &  $2.81\times 10^{-4}$  & $7.48\times 10^{-10}$ & $5.23\times 10^{0}$ & $6.30\times 10^{-5}$ \\
\end{tabular}
  \end{center}
  \caption{BLUP relative accuracy errors. For different parameters of
    the test, the condition numbers of the multilevel approach improve
    significantly over the traditional BLUP formulae.  The multilevel
    approach leads to high accuracy results. This is in contrast to
    the traditional BLUP formulae, which struggle to achieve
    reasonable accuracy.}
\label{numericalresults:table5}
\end{table*}

\subsection{Estimation}

In this section estimation results are presented for the Mat\'{e}rn
covariance matrix on high dimensional n-Sphere random locations by
solving multilevel log-likelihood
$\hat{\btheta} : =
\argmax_{\btheta}
\tilde{\ell}^{i}_{\bW}(\bZ^{i,k,d}_{W};\btheta)$,
where $\bZ^{i,k,d}_{W} := [\bW_t \T, \dots, \bW_i \T ] \T
\bZ^{d}_{k}$, for $i = t, t-1, \dots$. The observation data $\bZ^d_k$
is built from the n-Sphere $\bS^d_k$ for $d = 3,10$, $k = 6$ $(N =
64,000)$ and $k = 7$ $(N = 128,000)$. The covariance function is
Mat\'{e}rn for several values of $\nu$ and $\rho$. To test the
performance of the multilevel estimator, $M = 100$ realizations are
generated 

The optimization problem of the log-likelihood function
\eqref{Introduction:multilevelloglikelihoodreduced2} (and
\eqref{Introduction:multilevelloglikelihoodreduced3}) is solved using
a fmincon iteration search for the estimates $\hat \nu$ and $\hat
\rho$ from the optimization toolbox in MATLAB \cite{Matlab2019}. The
tolerance level is set to $10^{-6}$.  In Table
\ref{numericalresults:table2} the mean and standard deviation of the
Mat\'{e}rn covariance parameter estimates $\hat \nu$ and $\hat \rho$
are presented. The mean estimate $\bbE_M [\hat{\nu}]$ refers to the
mean of $M$ estimates $\hat{\nu}$ for the $M$ realizations of the
stochastic model. Similarly, $std_M [\hat{\nu}]$ refers to the
standard deviation of the $M$ realizations. For case (a) ($d = 3$) the
error mean and std is $\approx 1\%$. For case (b) ($d = 10$) the error
of the mean increase to $\approx 10 \%$. In general, as $i$ is reduced
from $t$ there is a tendency of a drop in the standard deviation
$std_M [\hat{\nu}]$ of the estimator $\hat \nu$. However, there is
also a tendency for the accuracy of the mean to degrade somewhat,
except for (a) $N = 128,000$, $i = 12 \rightarrow 11$.

\begin{table*}[htpb]
\caption{Estimation of parameters $\hat \nu$ and $\hat \rho$ with:
  Total Degree polynomial index set $\mcQ^d_w$, kD tree, and n-Sphere
  with for $d = 3$ and $d = 10$.  The observation data $\bZ^d_k$ are
  formed from the Mat\'{e}rn covariance function. The number of
  realizations of the Gaussian random field model is set to $M =
  100$. Several cases are tested and are given by the individual tables
  (a) and (b).  The first to fourth columns are self-explanatory. The
  fifth column is the mean error of $\hat \nu$ with $M$
  realization. The sixth column is the mean error of $\hat \rho$. The
  last two columns are the standard deviation of $M$ realizations of
  the parameters $\hat \nu$ and $\hat \rho$.}
\footnotesize
\begin{center}
(a) TD, kD tree, n-Sphere, $d = 3$, $M = 100$, $\nu =
  3/4$, $\rho = 1/6$, $\tau = 5 \times 10^{-2}$
\begin{tabular}{ r r r r r r r r r r}
\multicolumn{1}{c}{$N$} & 
\multicolumn{1}{c}{$w$} &
\multicolumn{1}{c}{$t$} &
\multicolumn{1}{c}{$i$} & 
\multicolumn{1}{c}{$\bbE_M [\hat{\nu} - \nu]$} &
\multicolumn{1}{c}{$\bbE_M [\hat{\rho} - \rho]$} &
\multicolumn{1}{c}{$std_M [\hat{\nu}]$} &
\multicolumn{1}{c}{$std_M [\hat{\rho}]$} 
 \\ 
 \hline
64000 & 3 & 11 & 11 & -1.92e-04 &  4.52e-04 & 1.36e-02 & 8.17e-03 \\ 
64000 & 3 & 11 & 10 &  1.17e-03 & -5.90e-04 & 7.08e-03 & 4.04e-03 \\ 
128000 & 3 & 12 & 12 & -2.51e-03 & 1.81e-03 & 8.54e-03 & 6.11e-03 \\ 
128000 & 3 & 12 & 11 & -6.90e-04 & 5.02e-04 & 4.17e-03 & 2.84e-03 \\ 
\end{tabular}
\\
\bigskip
(b) TD, kD tree, n-Sphere, $d = 10$, $M = 100$, $\nu =
  3/4$, $\rho = 3/4$, $\tau = 1 \times 10^{-5}$
\begin{tabular}{ r r r r r r r r r r}
\multicolumn{1}{c}{$N$} & 
\multicolumn{1}{c}{$w$} &
\multicolumn{1}{c}{$t$} &
\multicolumn{1}{c}{$i$} & 
\multicolumn{1}{c}{$\bbE_M [\hat{\nu} - \nu]$} &
\multicolumn{1}{c}{$\bbE_M [\hat{\rho} - \rho]$} &
\multicolumn{1}{c}{$std_M [\hat{\nu}]$} &
\multicolumn{1}{c}{$std_M [\hat{\rho}]$} 
 \\ 
 \hline
64000 & 4 & 5 & 5 &  8.70e-03 & -1.12e-02 & 1.55e-02 & 1.85e-02 \\ 
64000 & 4 & 5 & 4 & -9.31e-02 &  8.02e-02 & 1.67e-02 & 1.97e-02 \\
128000 & 4 & 6 & 6 & -6.36e-03 & 5.51e-03 & 2.10e-02 & 1.72e-02 \\
128000 & 4 & 6 & 5 & -7.18e-02 & 6.27e-02 & 1.32e-02 & 1.46e-02 \\ 
\end{tabular}
\end{center}
\label{numericalresults:table2}
\end{table*}

\section{Prediction}

In this section the computational performance of the multilevel solver
is analyzed. Given a fixed Mat\'{e}rn parameters $(\nu,\rho)$ the BLUP
vectors $\hat \bgamma$ and $\hat \beta$ are computed. This involves
solving the system of equations $\bP^{-1}_{\bW} \bC_{\bW}(\btheta)
\bgamma_{\bW} = \bP^{-1}_{\bW} \bZ_{\bW}$ and $\hat \bbeta = (\bX\T
\bX)^{-1}$ $\bX\T(\bZ - \bC \hat \bgamma)$. Results for computing
$\hat \bgamma$ and $\hat \bbeta$ for the hypercube data set with $d =
3$ dimensions, kD tree, and the Total Degree index set $\mcQ^d_w$ are
shown in Table \ref{numericalresults:table3}. The Mat\'{e}rn
covariance coefficients $\btheta = (\nu,\rho)$ are set to (3/4,1). The
relative error of the residual of PCG method for the unpreconditioned
system is set to $\varepsilon = 10^{-3}$. The KIFMM is set to high
accuracy.

For computing the matrix vector products of the PCG iterations, the
computational break even point of the KIFMM solver is reached for $N
\approx 2,500$ compared to using the direct approach (with CPU and
GPU). The increase in computational complexity is linear with respect
to $N$. Thus all the matrix vector products for the PCG iterations are
calculated using the KIFMM.  The preconditioner $\bP_{\bW}$ is built
using a combination of the GPU and CPU. This leads to a quadratic
increase in computational cost with respect to the number of
observations $N$. However, due to the high efficiency of the
implementation and $p = 120$, the break even point for the use of the
KIFMM solver is not reached, even for $N = 512,000$ observation
points.  From Table \ref{numericalresults:table3} observe that
condition number of the covariance matrix $\bC$ is much larger
compared to $\bC_{\bW}$. This is already a good indication that
solving the prediction problem will be more efficient using the
multilevel approach.

The number of iterations needed to reach the same accuracy for both
approaches are significantly better with the multilevel approach
i.e. $\approx 70$ times less iterations. However, the computation of
$\bbeta$ with the single level method requires solving $p= 121$
matrix inversions of $\bC$. This is in contrast with a single matrix
inversion of $\bC_{\bW}$ with the multilevel method. In practice, we
did not solve all $p$ matrix inversions for the single level
approach, but measure the time required to compute a single matrix
inversion and multiplied it 121 to obtain the estimated time
complexity.  For $N = 64,000$ observations we observe efficiencies of
$\approx 7,000$ compared to the single level iterative approach.
\begin{table*}[htbp]
  \setlength{\tabcolsep}{3.5pt}
  \caption{Numerical results for computing $\bP^{-1}_{\bW}
    \bC_{\bW}(\btheta) \bgamma_{\bW} = \bP^{-1}_{\bW} \bZ_{\bW}$ and
    $\hat \bbeta = (\bX\T \bX)^{-1} \bX\T(\bZ - \bC \hat \bgamma)$ for
    the hypercube data set with $d = 3$ and the Total Degree index set
    $Q^d_w$. The Mat\'{e}rn covariance coefficients $\btheta =
    (\nu,\rho)$ are set to (3/4,1). The relative error of the residual
    of PCG method for the \emph{unpreconditioned system} is set to
    $\varepsilon = 10^{-3}$. The KIFMM is set to high accuracy. (a)
    The second column of is the condition number of the covariance
    matrix $\bC$, up to $N=64,000$ observations, and is compared with
    the third column which corresponds to the condition number of
    $\bC_W$. The third column is the number of CG iterations needed
    for convergence for $10^{-3}$ residual accuracy. The fourth column
    is the number of iterations need to achieve the residual error
    $10^{-3}$ for the unpreconditioned system with the preconditioner
    $\bP_{\bW}$.  (b) The sixth column corresponds to the wall clock
    times in seconds for the preconditioner computation. The PCG
    iteration wall clock timings for $\bC_{\bW}$, by using a KIFMM,
    are given in the seventh column. The eighth column is the total
    time to compute $\bgamma_{\bW}$, $\bbeta$ and the multilevel basis
    construction. The eighth column is the computational efficiency for
    computing $\bgamma_{\bW}$ vs $\bC^{-1} \bZ$ to same residual
    accuracy with respect to the number of iterations. The last column
    is the estimated efficiency of computing $\hat \bgamma$ and $\hat
    \bbeta$ with the multilevel BLUP compared to the single level
    approach, equation \eqref{Spatial}, to approximately the same
    accuracy using a CG iteration with the KIFMM. We observe the
    significant speed ups ($\approx 7,000$ for $N = 64,000$) for
    calculating the BLUP by using the multilevel approach.  }
  \centering \footnotesize
  \medskip
$\btheta = (3/4,1)$, $d = 3$, $w = 7$ ($p = 120$) \\
\begin{tabular} { r c c r r r r r c r}    
  \multicolumn{1}{c}{$N$} 
  & $\kappa(\bC)$ & $\kappa(\bC_{\bW})$ & itr($\bC$) & itr($\bC_{\bW}$)
  & $\bP_{\bW}$ (s) & Itr (s) & Total (s) & Eff$_{\bgamma}$ &
  \multicolumn{1}{c}{Eff$_{\bgamma,\bbeta}$} \\
  \hline
  8,000   &  $3.2 \times 10^{7}$ & $1.8 \times 10^{4}$ &  1,985  &  52  &       4   &      29  &     38 &  38 &   3,600  \\
  16,000  &  $1.1 \times 10^{8}$ & $6.0 \times 10^{4}$ &  3,511  &  67  &      13   &      98  &    118 &  52 &   5,000  \\
  32,000  &  $5.6 \times 10^{8}$ & $3.1 \times 10^{5}$ &  8,259  & 116  &      45   &     260  &    317 &  71 &   7,250  \\
  64,000  &  $1.8 \times 10^{9}$ & $9.5 \times 10^{5}$ &  12,680 & 165  &     178   &     798  &    997 &  76 &   7,380  \\
  128,000 &                   - &                  -  &        -& 308  &     713   &   3,934  &  4,687 &   - &    - \\
  256,000 &                   - &                  -  &        -& 292  &   2,837   &   5,745  &  8,663 &   - &    - \\
  512,000 &                   - &                  -  &        -& 484  &   11,392  &  20,637  & 32,202 &   - &    - \\
\end{tabular}
\label{numericalresults:table3}
\end{table*}
The multilevel approach is now tested on $d = 20,25$ dimensional
problems. Due to the high dimensionality of these problems, a fast
summation approach is not an option. The matrix-vector products of
each iteration are computed with the direct approach using the GPU and
CPU.  In Table \ref{numericalresults:table4}(a) the numerical results
for computing $\bgamma$ and $\bbeta$ for $d = 20$ and $\theta
=(5/4,10)$.  Compared to the single level iterative approach the
multilevel method is approximately 42,000 faster for $N = 64,000$
observations. Similar results are obtained shown in Table
\ref{numericalresults:table4}(b).  for $d = 25$ and $\theta
=(5/4,10)$.

\setlength{\tabcolsep}{3pt}

\begin{table*}[htbp]
  \caption{Computing BLUP for the n-sphere data set with $d = 20$ and
    $d = 25$ dimensions, TD index set, and Mat\'{e}rn covariance
    function without pre-conditioner. The residual accuracy is set to
    $\varepsilon = 10^{-3}$. Since the dimension is greater than 3,
    the matrix vector products are computed directly with the GPU and
    CPU.  The description of the columns of tables (a) and (b) are the
    same as for Table \ref{numericalresults:table3}. In addition,
    column 6 corresponds to the wall clock time for computing the
    multilevel basis. (a) Computational times for solving the
    prediction for $d = 20$ and $\theta = (5/4,10)$.  The growth in
    computational cost is slightly faster than quadratic due to the
    lack of fast summation method in higher dimensions. However,
    compared to the single level iterative approach it is
    approximately 42,000 faster for $N = 64,000$ observations. (b)
    BLUP prediction for $d = 25$ and $\theta = (5/4,10)$. The growth
    in computational cost is similar.  The efficiency of this method
    is about 2,840 times faster.}  \footnotesize
  \begin{center}
(a) $\btheta = (\nu,\rho) = (5/4,10)$, $d = 20$, $w = 3$ ($p =
1771$), No precond., Direct \\
\begin{tabular} { r c c c  c r r r r r}
  \multicolumn{1}{c}{$N$} & $\kappa (\bC)$ & $\kappa (\bC_{\bW})$ &
 itr($\bC$)
  & 
 itr($\bC_{\bW}$) &  MB(s) & Itr(s) & Total(s) &  Eff$_{\bgamma,\bbeta}$ \\
  \hline
 16,000  & $5  \times 10^{7}$ & 7  & 238 & 10  &  52  &      97   &    153 & 26,700   \\
 32,000  & $1 \times 10^{8}$  & 11 & 324 & 13  & 121  &     500   &    628 & 35,160   \\
 64,000  & $2 \times 10^{8}$  & 17 & 444 & 17  & 284  &   2,600   &  2,898 & 42,050 \\
128,000  &  -                & -  &  - & 22  & 628  &  13,494   &  14,153 & -  \\
\end{tabular}\\
\bigskip
(b) $\btheta = (\nu,\rho) = (3/4,10)$, $d = 25$, $w = 2$ ($p = 351$), No precond., Direct \\
\begin{tabular} { r c c c c r r r r}
  \multicolumn{1}{c}{$N$} & $\kappa (\bC)$ & $\kappa (\bC_{\bW})$
  & itr($\bC$)
  & itr($\bC_{\bW}$) & MB(s) & Itr(s) & Total(s) & Eff$_{\bgamma,\bbeta}$ \\
  \hline
 16,000  & $2  \times 10^{6}$  &   7  & 86  & 12 &   5  &         116   &    122 & 2,400   \\
 32,000  &  $4   \times 10^{6}$ &  12 & 109 & 15 &  13  &         582   &    599 & 2,490 \\
 64,000  &  $9  \times 10^{6}$ &   21 & 147  & 18 &  30  &       2,788  &  2,821 & 2,840 \\
 128,000  &  -                  &  - & -  & 25 &  79  &      15,557   & 15,641 &  - \\
 256,000  &  -                  &  - & -  & 33 & 157  &      83,163   & 83,337 &  - \\
\end{tabular}\\
\end{center}
\label{numericalresults:table4}
\end{table*}

\section{Conclusions}

In this paper a multilevel method is developed that scales well with
high dimensions for solving the spatial BLUP. A multilevel basis is
constructed from a kD-tree and for the choice of Total Degree
polynomial basis $\mcQ^d_w$.  The approach described in the paper has
the following characteristics and advantages:

\begin{inparaenum}[i)]

\item The multilevel method is numerically stable. Hard estimation and
  prediction of large dimensional problems are now feasible.
\item The method is efficiently implemented by using a combination of
  MATLAB, c++ software packages and dynamic libraries.
\item Sub-exponential decay of multilevel covariance matrix
  $\bC_{\bW}$ is proven based on complex analytic extensions.
\item Numerical results of up to 25 dimensional problems. These
  problems are difficult to solve with traditional methods due to the
  large condition numbers, but feasible with the multilevel method.
\item The multilevel prediction approach is proven to be \emph{exact},
  in the sense that single level and multilevel prediction
  formulations are shown to be equivalent.
\item The efficiency of this approach will be further improved as high
  dimensional fast summation methods are developed.
\item An A-posteriori scheme and estimates for constructing the sparse
  covariance matrix $\tilde \bC$ will be developed in a future
  paper. This will be possible with the error bounds for the entries
  of $\bC$ derived in this paper since all the constants can be
  estimated.
\end{inparaenum}

\textbf{Acknowledgments:} I appreciate the help and advice from George
Biros and Lexing Ying for setting up the KIFMM packages. In addition,
I am grateful to the intense and very illuminating discussions with
Michael Stein and Mihai Anitesc.  I also appreciate the support that
King Abdullah University of Science and Technology has provided to
this project.

\begin{appendix}

\section[Polynomial Interpolation]{Polynomial Interpolation}
\label{PolynomialAppendix}

In this section we provide some background on polynomial interpolation
in high dimensions. This will be critical to estimate the decay rates
of the entries of the multilevel covariance matrix for high
dimensional problems.

The decay of the coefficients will directly depend on the analytic
properties of the covariance function. The traditional error estimates
of polynomial interpolation are based on multi-variate $m^{th}$ order
derivatives. However, for many cases, such as the Mat\'{e}rn
covariance function, the derivatives are too complex or expensive to
manipulate for even a moderate number of dimensions. This motivates
the study of polynomial numerical approximations based on complex
analytic extensions, which are much better suited for high dimensions.
Much of the discussion that follows has it roots in the field of
uncertainty quantification and high dimensional interpolation
\cite{nobile2008a,Castrillon2016,Griebel2016}
for partial differential
equations.

Consider the problem of approximating a function $v: \Gamma^{d}
\rightarrow \R$ on the domain $\Gamma^{d}$.  Without loss of
generality let $\Gamma \equiv [-1, 1]$ and $\Gamma^{d} \equiv \prod_{n =
  1}^{d} \Gamma$. Suppose that $\mcG \subset \Gamma^{d}$, then define
the following spaces
\begin{equation*}
\begin{split}
  &
L^q(\mcG) := \{ v(\by)\, : \, \int_{\mcG} v(\by)^q \text{d}
\by < \infty  \}
\,\,\,
\mbox{and} \,\,\,\,
L^{\infty}(\mcG) := \{ v(\by)\, : \, \sup_{\by \in \mcG} \lvert v(\by) \rvert
< \infty  \}.
\end{split}
\end{equation*}

Suppose that $\mcP_{ q}(\Gamma):=\text{\rm span}\{y^k,\,k=0,\dots,q\}$
i.e. the space of polynomials of degree at most $q$. Let $\mcI^{m} :
C^{0}(\Gamma) \rightarrow \mcP_{m-1}(\Gamma)$ be the univariate
Lagrange interpolant
$\mcI_{m}(v(\by)):= \sum_{k=1}^{m}v(y^{(k)})l_{m,k}(y^{(k)})$,
where $y^{(1)}, \dots, y^{(m)}$ is a set of distinct knots on $\Gamma$
and $\{ l_{n,k} \}_{k=0}^{m}$ is a Lagrange basis of the space
$\mcP_{m-1}(\Gamma)$. The variable $m \in \Nset$
corresponds to the order of approximation of the
Lagrange interpolant. However, for the case of the zero order
interpolation $m = 0$ corresponds to $\mcI_{0} = 0$.

\begin{remark}
For high dimensional interpolation the particular set of points
$y^{(1)}, \dots,$ $y^{(m)}$ that we will use is the Clenshaw-Curtis
abscissas.  This is further discussed in this section. However, for
now, we assume that the points are only distinct.
  \end{remark}

For $m \geq 1$ let
$\Delta_{m}
:= \mcI_{m}-\mcI_{m-1}$.
From the difference operator $\Delta_{m}$ we can readily observe that
$\mcI_{m} = \sum_{k=1}^{m} \Delta_{k}$, which is reminiscent of multi
resolution wavelet decompositions. The idea is to represent
multivariate approximation as a summation of the difference operators.

Consider the multi-index tupple $\bm = (m_1,\dots,m_d)$, where $\bm
\in \Nset^{d}$, and form the tensor product operator
$\mcS_{w,d}: \Gamma \rightarrow \R$ as
\begin{equation}
  \mcS_{w,d}
      [v(\by)]
      :
      =
 \sum_{\bm \in \bbNset^{d}: \sum_{i=1}^{d} m_i - 1  \leq w } \;\;
 \bigotimes_{n=1}^{d} {\Delta^{n}_{m_n}}(v(\by)).
\label{errorestimates:SG}
\end{equation}
Note that by ${\Delta^{n}_{m_n}}(v(\by))$ we mean that the difference
operator ${\Delta_{m_n}}$ is applied along the $n^{th}$ dimension in
$\Gamma$.

Let $C^{0}(\Gamma^d; \R) : = \{ v: \Gamma^d \rightarrow \R\,\,$ is
continuous on $\Gamma^d$ and $\max_{\by\in \Gamma^d} \lvert v(\by)
\rvert < \infty \}$.  From Proposition 1 in \cite{Back2011} it is
shown that for any $v \in C^0(\Gamma^d;\R)$, we have $\mcS_{w,d}[v]\in
\mcQ^{d}_{w}$.  Moreover, $\mcS_{w,d}[v] = v$, for all $v \in
\mcQ^{d}_{w}$. The key observation to take away is that the operator
$\mcS_{w,d}[v]$ is \textit{exact} in the space of polynomials
$\mcQ^{d}_{w}$. This will be useful in connecting the Lagrange
interpolant with Chebyshev polynomials.

Let $T_k:\Gamma \rightarrow \R$, $k = 0, 1, \dots$, be a Chebyshev
polynomial over $\Gamma$, which are defined recursively as follows:
$T_0(y) = 1$, $T_1(y) = y$, $\dots$, $T_{k+1}(y) = 2yT_{k}(y) -
T_{k-1}(y)$, $\dots$, where $y \in \Gamma$. Chebyshev polynomials are
well suited for the approximation of functions with analytic
extensions on a complex region bounded by a Bernstein ellipse. They
bypassing the need of using derivative information and sharp bounds on
the error are readily available. Suppose that $\sigma > 0$ and denote
by
\begin{equation*}
\begin{split}
  \mcE_{\sigma} &:= \Big\{
  z \in \bbC, \sigma \geq
\delta \geq 0 :\,\Real{z} = \frac{e^{\delta} + e^{-\delta}
}{2}cos(\theta), 
\Imag{z} = \frac{e^{\delta} 
  - e^{-\delta}}{2}sin(\theta), \\
&\theta \in [0,2\pi)
  \Big\}
\end{split}
  \end{equation*}
as the region bounded by a Bernstein ellipse (see Figure
\ref{erroranalysis:sparsegrid:polyellipse}).  The following theorem is
based on complex analytic extensions on $\mcE_{\sigma}$ and provides a
control for the Chebyshev polynomial approximation.

\begin{theorem}
Suppose that for $u:\Gamma \rightarrow \R$ there exists an analytic
extension on $\mcE_{\sigma}$. If $|u| \leq M < \infty$ on
$\mcE_{\sigma}$ then there exists a sequence of coefficients
$\lvert\alpha_k\rvert \leq M / e^{k\sigma}$ such that $u \equiv \alpha_0 +
2\sum_{k = 1}^{\infty} \alpha_{k} T_{k}$ on $\mcE_{\sigma}$. Moreover,
if $y \in \Gamma$ then
$\vert q(y) - \alpha_0  - 2\sum_{k = 1}^{n} \alpha_{k} T_{k}(y) \rvert
\leq 
\frac{2M}{e^{\sigma} - 1} e^{-n \sigma}$.
\label{errorestimates:theorem}
\end{theorem}
\begin{proof}
See Theorem 2.25 in \cite{Khoromskij2018}
\end{proof}

\begin{figure}[htb]\begin{center}
\begin{tikzpicture}
    \begin{scope}[font=\scriptsize]

    \fill [pattern=north west lines, pattern color=blue,semitransparent]
      (0,0) ellipse (2 and 1);

    \draw [->] (-2.5, 0) -- (2.5, 0) node [below left]  {$$};
    \draw [->] (0,-1.5) -- (0,1.5) node [below left] {$i$};
    \draw (1,-3pt) -- (1,3pt)   node [above] {$1$};
    \draw (-1,-3pt) -- (-1,3pt) node [above] {$-1$};
    \end{scope}
    
    \node [below right] at (-2.5,1.25) {$\mcE_{\sigma}$};

    \node [] at (0.75,1.25) {$\frac{e^{
          \sigma} - e^{- \sigma}}{2}$};

    \node [] at (2.75,0.25) {$\frac{e^{
      \sigma} + e^{- \sigma}}{2}$}; 
    
\end{tikzpicture}
\end{center}
\caption{Complex region bounded by the Bernstein ellipse.}
\label{erroranalysis:sparsegrid:polyellipse}
\end{figure}

We can now connect the error due to the Lagrange interpolation with
Chebyshev expansions. It is known that if $u \in C(\Gamma,\R)$ then
$\|(I - \mcI_{m})u\|_{L^{\infty}(\Gamma)}$ $\leq (1 + \Lambda_{m})
\min_{h \in \mcP_{m-1}} \| u - h \|_{L^{\infty}(\Gamma)}$, where
$\Lambda_{m}$ is the Lebesgue constant (See Lemma 7 in
\cite{babusk_nobile_temp_10}). Note that $I:C^{d}(\Xi;\R) \rightarrow
C^{d}(\Xi;\R)$ refers to the identity operator and the domain $\Xi$ is
taken from context. For the previous case $\Xi = \Gamma$.  Bounds on
$\Lambda_{m}$ are known in the context of the location of the knots
$y^{(1)}, \dots, y^{(m)} \in \Gamma$. In this article we restrict our
attention to Clenshaw-Curtis abscissas $y^{(j)} = -\cos \left(
\frac{\pi(j-1)}{m - 1} \right),\,\, j = 1,\dots, m$ and $\Lambda_m$ is
bounded by $2\pi^{-1}(\log{(m-1)} + 1) \leq 2m - 1$ (see
\cite{babusk_nobile_temp_10}).  Since the interpolation operator
$\mcI_{m}$ is exact on $\mcP_{m - 1}$, then if $u:\Gamma \rightarrow
\R$ has an analytic extension in $\mcE_{\sigma}$ we have from Theorem
\ref{errorestimates:theorem} (following a similar approach as in
\cite{babusk_nobile_temp_10}) that $\|(I -
\mcI_{m})u\|_{L^{\infty}(\Gamma_n)} \leq (1 + \Lambda_{m})
\frac{2M}{e^{\sigma} - 1} e^{-\sigma (m-1)} \leq 2 C(M,\sigma) m
e^{-\sigma (m-1)}$,
\begin{comment}
This gives us the following result
\begin{theorem} Suppose that $0< \delta < 1$, $\hat
  \sigma := \sigma (1 - \delta)$, and $\phi(\bx,\by;\btheta) \in
  C^{0}(\Gamma^{d} \times \Gamma^{d};\R)$ can be analytically extended
  on $\mcE^{d}_{\sigma} \times \mcE^{d}_{\sigma}$ and is bounded by
  $\tilde M(\phi)$. Let $\mcP^{ p}(\mathbb{S})^{\perp}$ be the
  subspace in $\R^{N}$ generated by the index set $\mcQ^{d}_{w}$ for
  some $w \in \bbN_{+}$. For $i,j = 0,\dots,t$ consider any
  multilevel vector $\bpsi^{i}_m \in \mcP^{p}(\mathbb{S})^{\perp}$,
  with $n_m$ non-zero entries, from the cell $B^{i}_{m} \in \mcB^{i}$
  and any multilevel vector $\bpsi^{j}_{q} \in \mcP^{
    p}(\mathbb{S})^{\perp}$, with $n_q$ non-zero entries, from the
  cell $B^{j}_{q} \in \mcB^{j}$. If $p(d,w) \geq \left(\frac{2
    d}{\kappa(d)}\right)^{d}$ then
\begin{equation*}
\begin{split}
\lvert
\sum_{r = 1}^{N} 
\sum_{h = 1}^{N} 
\phi(\bx_r,\by_h; \btheta) 
\bpsi^i_m[h] \bpsi^j_q[r] 
\rvert
&\leq
\sqrt{n_mn_q}
\left( \frac{
  C(\tilde M,\sigma)^{d} e^{d - \sigma(1 - \delta) + 1} \hat \sigma d
}
 {
   (\sigma \delta)^{d}} \right)^2
 \\
 &
 \left( \frac{e^{\hat \sigma}}{1 - e^{-\hat \sigma}} \right)^{2d}
 \exp \left(-\frac{2d}{e} \hat \sigma  p^{\frac{1}{d}}
 \right) p^{2\left(\frac{d-1}{d}\right)}.
\end{split}
\end{equation*}
\end{theorem}
\end{comment}
where 
$C(M,\sigma_n) := \frac{2M}{(e^{ \sigma} - 1)}$. We then
conclude that for all $k = 1,\dots, m$
\begin{equation}
\begin{split}
\| \Delta_{k}(u) \|_{L^{\infty}(\Gamma)} 
&=
\|
\mcI^{m}(u) - \mcI^{m-1}(u)
\|_{L^{\infty}(\Gamma)} 
\leq
\|(I - \mcI_{m})u\|_{L^{\infty}(\Gamma)} \\
&+
\|(I - \mcI_{m-1})u\|_{L^{\infty}(\Gamma)} 
\leq
e^{2\sigma}C(M,\sigma) m e^{-\sigma m}.
\end{split}
\label{interpolation:eqn1}
\end{equation}
Let $\mcE_{\sigma,n} \subset \bbC^{d}$ a complex region bounded by a
Bernstein ellipse such that the restriction on $\Gamma_{d}$ is along
the $n^{th}$ dimension and form the polyellipse $\mcE^{d}_{\sigma}:=
\prod_{n=1}^{d} \mcE_{\sigma,n}$.  Suppose that $v:\mcE^{d}_{\sigma}
\rightarrow \bbC$ is analytic on $\mcE^{d}_{\sigma}$ and let
$\tilde{M}(v) := \max_{\bz \in \mcE^{d}_{\sigma}} \vert v(\bz) \rvert$.

Note we refer to $\mcI^{n}_{m}$ as the Lagrange operator of order $m$
along the $n^{th}$ dimension and similarly $\mcP^{n}_{m-1}$ is the
space of the span of univariate polynomials up to degree $m-1$ along
the $n^{th}$ dimension.  Form the tensor product $\bI^{d}_{m} :=
\mcI^{1}_{m} \times \dots \times \mcI^{d}_{m}$, thus $\bI:C(\Gamma,\R)
\rightarrow \bbP$ where $\bbP := \mcP^{1}_{m-1} \times \dots \times
\mcP^{d}_{m-1}$. From Theorem 2.27 in \cite{Khoromskij2018} we can
conclude that for a finite dimension $d$, as $m \rightarrow \infty$
then $\bI^{d}_{m}[v] \rightarrow v$.

Applying equation \eqref{interpolation:eqn1} to equation
\eqref{errorestimates:SG} we have that
\begin{equation}
\begin{split}
& \| (I - \mcS_{w,d})
 v(\by)
 \|_{L^{\infty}(\Gamma^{d})}
 \leq
 \left\| \sum_{\bm \in \bbNset^{d}: \sum_{i=1}^{d} m_i - 1 > w } \;\;
 \bigotimes_{n=1}^{d} {\Delta^{n}_{m_n}}(v(\by))\right\|_{L^{\infty}(\Gamma^d)} \\
 &\leq
 \sum_{\bm \in \bbNset^{d}: \sum_{i=1}^{d} m_i - 1 > w } 
 \bigotimes_{n=1}^{d} \|{\Delta^{n}_{m_n}}(v(\by))\|_{L^{\infty}(\Gamma^d)} \\
 &\leq
 \sum_{\bm \in \bbNset^{d}: \sum_{i=1}^{d} m_i - 1 > w }
 e^{2d} C(M,\sigma)^{d} 
 \left( \prod_{n=1}^{d} m_n\right) \exp{\left( -\sum_{n=1}^{d}
   \sigma m_{n} \right)}.
\end{split}
\label{interpolation:eqn2}
\end{equation}

By applying Theorem 2.10 and Corollary 2.11 in \cite{Griebel2016} if
$ w \geq  d$ and $p( d, w) \geq
\left(\frac{2  d}{\kappa( d)}\right)^{ d}$, where
$\kappa( d) := \sqrt[\leftroot{-2}\uproot{2}  d]{
  d!} >  d/e$ (Sterling approximation), then for any $\hat
\sigma \in \R_{+}$
\begin{equation}
\begin{split}
 & \sum_{\bk \in \bbNset^{ d}_{0}: \sum_{i=1}^{ d} k_i  >  w }
 \exp{\left( -\sum_{n=1}^{ d} \hat \sigma
   k_{n} \right)}
 \leq
 \sum_{\bk \in \bbNset^{d}_{0}: \hat \sigma \sum_{i=1}^{ d} k_i  \geq  w \hat \sigma  }
 \exp{\left( -\sum_{n=1}^{ d}
   \hat \sigma k_{n} \right)} \\
 &\leq
 \hat \sigma  d e
 \left( \frac{e^{\hat \sigma}}{1 - e^{-\hat \sigma}} \right)^{ d}
 \exp \left(-\frac{ d}{e} \hat \sigma  p^{\frac{1}{ d}}
 \right) p^{\frac{ d-1}{ d}}.
\end{split}
\label{interpolation:eqn3}
\end{equation}
where $\bk \in \bbNset^{d}_{0}$ and $\bk:=(k_1,\dots,k_d)$.

Following the same approach as in \cite{Griebel2016} observe that for
$0 < \delta < 1$ we can obtain a bounded constant $c_{n,\delta} \leq
(e\sigma \delta)^{-1}$ such that $m_n \exp(-\sigma m_n) \leq (e\sigma
\delta)^{-1}$ $\exp(-\sigma m_n (1 - \delta))$. Set $\hat \sigma :=
\sigma (1 - \delta)$ and by combining equations
\eqref{interpolation:eqn2} and \eqref{interpolation:eqn3} we have
proven the following result.

\begin{lemma} Suppose that $0< \delta < 1$, $\hat
  \sigma := \sigma (1 - \delta)$, and $p(d,w) \geq \left(\frac{2
    d}{\kappa(d)}\right)^{d}$ then
  \[
  \begin{split}
  \| (I - \mcS_{w,d})
 v(\by)
 \|_{L^{\infty}(\Gamma^{d})}
 &\leq 
 \frac{C(\tilde M,\sigma)^d e^{d - \sigma(1 - \delta) + 1} \hat \sigma d }
 {
(\sigma \delta)^{d}}
 \left( \frac{e^{\hat \sigma}}{1 - e^{-\hat \sigma}} \right)^{d} \\
 &\exp \left(-\frac{d}{e} \hat \sigma  p^{\frac{1}{d}}
 \right) p^{\frac{d-1}{d}}.
 \end{split}
 \]
\label{interpolation:lemma1}
\end{lemma}

\begin{remark}
The restriction $p(d,w) \geq \left(\frac{2
  d}{\kappa(d)}\right)^{d}$ is not strict and can be relaxed such that
sub-exponential convergence is still obtained.  We refer the reader to
the bound of the Gamma function in Lemma 2.5 (\cite{Griebel2016}) and
it's application in the proofs of Theorem 2.10 and Corollary 2.11.
\label{interpolation:remark1}
\end{remark}
   \section{Experimental setup}
\label{experimental}

\begin{enumerate}

\item {\bf Matlab, C/C++ and MKL:} The binary tree, multilevel basis
  construction, formation of the sparse matrix $\tilde \bC_{\bW}$,
  estimation and prediction components are written and executed on
  Matlab \cite{Matlab2019}. However, the computational bottlenecks are
  executed by C/C++ software packages, Intel MKL \cite{intelmkl}, and
  the highly optimized BLAS and LAPACK packages contained in
  MATLAB. The C/C++ interfaces to matlab are constructed as dynamic
  shared libraries.

\item {\bf Direct and fast summation:} The matlab code estimates the
  computational cost between the direct and fast summation methods and
  chooses the most efficient approach.  For the direct method a
  combination of Graphic Processing Unit (GPU) and MKL intel libraries
  are used. For the fast summation method the KIFMM ($d = 3$) c++ code
  is used.  The KIFMM is modified to include a Hermite interpolant
  approximation of the Mat\'{e}rn covariance function, which is
  implemented with the intel MKL package \cite{intelmkl} (see
  \cite{Castrillon2015} for details).

\item {\bf Dynamic shared libraries:} These are produced with the GNU
  gcc/g++ packages. These libraries implement the Hermite interpolant
  with the intel MKL package (about 10 times faster than Matlab
  Mat\'{e}rn interpolant) and link the MATLAB code to the KIFMM.

\item {\bf Cholesky and determinant computation:} The Suite Sparse
  4.2.1 package
  (\cite{Chen2008,Davis2009,Davis2005,Davis2001,Davis1999}) is used
  for the determinant computation of the sparse matrix $\tilde
  \bC_{\bW}(\btheta)$.

\end{enumerate}

The code is tested on a single CPU (4 core Intel i7-3770 CPU @
3.40GHz.), one Nvidia 970 GTX GPU, with Linux Ubuntu 18.04 and 32 GB
memory. In addition, the Boston University Shared Computing Cluster
was used to generate test data.  To test the effectiveness of the
Multilevel solver the following data sets are generated:
\begin{enumerate}

\item {\bf Random n-sphere data set:} The set of nested random
  observation $\bS_{0}^{d} \subset \dots \subset \bS_{9}^{d}$ vary
  from 1,000, 2000, 4000 to 256,000 knots generated on the n-sphere
  $\bS_{d-1} := \{\bx \in \R^{d}\,\,:\,\,\|\bx\|_{2} = 1 \}$.

\item {\bf Random hypercube data set:} The set of random observation
  locations $\bC_{0}^{d},$ $\dots, \bC_{10}^{d}$ vary from 1,000, 2000,
  4,000 to 512,000 knots generated on the hypercube $[-1,1]^{d}$ for
  $d$ dimensions.  The observations locations are also nested,
  i.e. $\bC_{0}^{d} \subset \dots \subset \bC_{10}^{d}$.
  
\item {\bf Normal test data set} The set of observations values
  $\bZ^d_{0}$, $\bZ^d_{1}$, \dots $\bZ^d_{5}$ are formed from the
  Gaussian random field model \eqref{Introduction:noisemodel} for
  1,000, 2,000, $\dots$ $256,000$ observation locations. The data set
  $\bZ^d_{n}$ is generated from the set of nodes $\bS^{d}_{n}$, with
  the covariance parameters $(\nu,\rho)$ and the corresponding set of
  monomials $\mcQ^d_w$. The Boston University Shared Computing Cluster
  was used to generate the normal test data.

\end{enumerate}

\begin{remark}
 \corb{All the timings for the numerical tests are given in wall clock
   times i.e. the actual time that is needed to solve a problem. This
   is to distinguish it from CPU time, which can be significantly
   smaller and may not accurately reflect the real-world time taken
   for solving a problem.}
  \end{remark}
   \section{Proofs}

\textbf{Proof of \cref{Multilevelapproach:theo1}}

The proof is immediate.

  \medskip \noindent \textbf{Proof of \cref{MultilevelREML:lemma1}}

Starting at the finest level $t$, for each cell $B^{t}_k \in \mcB^{t}$
there is at most $p$ multilevel vectors.  Since there is at most
$2^t$ cells then there is at most $2^{t} p$ multilevel vectors.

Now, for each pair of left and right (siblings) cells at level $t$ the
parent cell at level $t-1$ will have at most $2 p$ scaling
functions. Thus at most $p$ multilevel vectors and $p$ scaling
vectors are obtained that are to be used for the next level. Now, the
rest of the cells at level $t$ are leafs and will have at most $p$
multilevel vectors and $p$ scaling vectors that are to be used for
the next level. Since there is at most $2^{t-1}$ cells at level $t-1$,
there is at most $2^{t-1} p$ multilevel vectors. Now, follow an
inductive argument until $q = 0$ and the proof is done.

\medskip \noindent \textbf{Proof of \cref{MultilevelREML:lemma2}}

For any leaf cell at the bottom of the tree (level $t$)
  there is at most $2 p$ observations.
cell has at most $4 p$ observations, thus the associated multilevel
  vectors has $4p$ non zero entries. By induction at any level $l$ the
  number of nonzero entries is at most $2^{t-q+1} p$.  Now for any
  leaf cell at any other level $l < t$ the number of nonzero entries
  is at most $2 p$. Following an inductive argument the result is
  obtained.

  \medskip \noindent \textbf{Proof of \cref{MultilevelREML:prop1}}
  
Let us look at the cost of computing all the 
interactions between any two cells $B^{i}_k \in \mcB^{i}$ and
$B^{j}_l \in \mcB^{j}$. Without loss of generality assume that $i
\leq j$. For the cell $B^{l}_k$ there is at most $p$
multilevel vectors and from Lemma \ref{MultilevelREML:lemma2}
$2^{t-i+1} p$ non zero entries. Similarly for $B^{j}_l$.  All
the interactions $(\bpsi^{i}_{\tilde{k}})\T \bC(\btheta)
\bpsi^{j}_{\tilde{l}}$ now have to be computed, where
$\bpsi^{i}_{\tilde{k}} \in B^{i}_k$ and $\bpsi^{j}_{\tilde{l}} \in
B^{j}_l$.

The term $\bC(\btheta) \bpsi^{j}_{\tilde{l}}$ is computed using a FMM
with $2^{t-j+1} p$ sources and $2^{t-i+1} p$ targets at a cost of
$\mcO($ $(2^{t-j+1} p + 2^{t-i+1} \tilde p)^{\alpha})$.  Since there
is at most $p$ multilevel vectors in $B^{i}_k$ and $B^{j}_l$ then the
cost for computing all the interactions $(\bpsi^{i}_{\tilde{k}})\T
\bC(\btheta) \bpsi^{j}_{\tilde{l}}$ is $\mcO(p(2^{t-j+1} p + 2^{t-i+1}
p)^{\alpha} + 2^{t-i+1}p)$.

Now, at any level $i$ there is at most $2^{i}$ cells, thus the result
follows.

\medskip \noindent \textbf{Proof of \cref{Multilevelapproach:theo2}}

A simple extension of the proof in Proposition \ref{Multilevelapproach:theo1}.

\medskip \noindent \textbf{Proof of \cref{multilevelSpatial:lemma2}}

Immediate.

\medskip \noindent \textbf{Proof of \cref{errorestimates:theorem1}}

We first have that
\begin{equation*}
\begin{split}
&\sum_{k = 1}^{N} 
\sum_{l = 1}^{N} 
\phi(\bx_k,\by_l; \btheta) 
\bpsi^i_m[k] \bpsi^j_q[l] 
=
\sum_{k = 1}^{N} 
\sum_{l = 1}^{N}
\lim_{g \rightarrow \infty}
(\bI^d_{g}
\otimes
\bI^d_{g})[
\phi(\bx_k,\by_l; \btheta)] 
\bpsi^i_m[k] \bpsi^j_q[l] \\
&=
\sum_{k = 1}^{N} 
\sum_{l = 1}^{N}
(I_d - \mcS_{w,d}) 
\otimes
(I_d - \mcS_{w,d}) 
[\phi(\bx_k,\by_l; \btheta)] 
\bpsi^i_m[k] \bpsi^j_q[l].
\end{split}
\end{equation*}
The last equality follows from $\bpsi^{i}_m, \bpsi^{j}_{q} \in \mcP^{
  p}(\mathbb{S})^{\perp}$. We now have that
\begin{equation*}
\begin{split}
& \sum_{k = 1}^{N} 
\sum_{l = 1}^{N}
\|(I_d - \mcS_{w,d}) 
\otimes
(I_d - \mcS_{w,d})
[\phi(\bx_k,\by_l; \btheta)] \|_{L^{\infty}_{\rho}(\Gamma^d)}
\lvert \bpsi^i_m[k] \rvert \lvert \bpsi^j_q[l] \rvert
\\
&\leq 
\|(I_d - \mcS_{w,d}))[\phi(\bx_k,\by_l; \btheta)] \|^{2}_{L^{\infty}_{\rho}(\Gamma^d)}
\sum_{k = 1}^{N} 
\sum_{l = 1}^{N} 
\vert\bpsi^i_m[k]\rvert \lvert\bpsi^j_q[l]\rvert.
\end{split}
\end{equation*}
Since $\bpsi^i_m$ and $\bpsi^j_q$ are orthonormal then
\[
\sum_{k = 1}^{N} 
\sum_{l = 1}^{N}
\lvert \bpsi^i_m[k] \rvert \lvert \bpsi^j_q[l] \rvert
\leq \sqrt{n_m n_q}
\|\bpsi^i_m[k]\|_{l^2} \|\bpsi^j_q[l]\|_{l^2}=
\sqrt{n_m n_q}.
\]
From Lemma \ref{interpolation:lemma1} the result follows.

\medskip \noindent \textbf{Proof of \cref{errorestimates:theorem2}}

The polynomial function is an entire function. However, the function
$K_{\nu}(\vartheta)$ and $\vartheta^{\frac{1}{2}}$ are analytic for
all $\vartheta \in \bbC$ except at the branch cut $(-\infty,0]$.  Thus
  it is sufficient to check the analytic extension of $r(\btheta) =
  \Big( \sum_{k=1}^{d} \theta_{k} (x_k - y_k)^{2}
  \Big)^{\frac{1}{2}}$.  Let $z \in \bbC$ be the complex extension of
  $r \in \R$. More precisely, $z = \Big( \sum_{k=1}^{d} \theta_{k}
  z_k^{2} \Big)^{\frac{1}{2}}$, where $z_k \in \bbC$ is the complex
  extension of $(x_k - y_k)$.

Let $\vartheta = \sum_{k=1}^{d} \theta_{k} z_k^{2}$, then by taking
the appropriate branch $\Real z = r_{\vartheta}$ $\cos{(
  \theta_{\vartheta}/2)}$, where $r^2_{\vartheta} = (\Real
\vartheta)^2 + (\Imag \vartheta)^2$ and $\theta_{\vartheta} =
\tan^{-1} \frac{\Imag \vartheta}{\Real \vartheta}$. Due to the branch
cut at $(-\infty,0]$ we impose the restriction that $\Real \vartheta >
0$ as $x_k$ and $y_k$ are extended in the complex plane.  Consider any
two cells $B^{i}_{m} \in \mcB^i$ and $B^{j}_{q} \in \mcB^j$, at levels
$i$ and $j$ with the associated distance criterion constant
$\tau_{i,j}>0$. From Algorithms \ref{MLCM:algorithm3},
\ref{MLCM:algorithm4}, \ref{MLCM:algorithm5} \ref{MLCM:algorithm6},
for any observations $\bx^{*} \in B^{i}_{m}$ and $\by^{*} \in
B^{j}_{q}$ we have that $(x^*_k - y^*_k)^2 \geq \tau^2_{i,j}$ for $k =
1,\dots,d$.  For the rest of the discussion it is assumed that complex
extension is respect to each component $k = 1,\dots,d$ unless
otherwise specified.

Extend $\alpha_k \rightarrow \alpha_k + v_k$ and $\gamma_k
\rightarrow \gamma_k + w_k$ where $v_k:= v^R_k + iv^I_k$, $w_k:= w^R_k
+ iw^I_k$, and $v^R_k,v^I_k,w^R_k,w^I_k \in \R$. Let $\tilde x_k$ be
the extension of $x_k$ in the complex plane and similarly for
$\tilde y_k$.
It follows that $\tilde x^R_k := \Real \tilde x_k = \frac{1}{2}
(\alpha_k + 1 + v^R_k)a_k + b_k$, $\tilde x^I_k = \Imag \tilde x_k =
\frac{v^I_k}{2} a_k$, $y^R_k := \Real \tilde y_k = \frac{1}{2}
(\gamma_k + 1 + w^R_k)c_k + d_k$, and $y^I_k := \Imag \tilde y_k =
\frac{w^I_k}{2} c_k$.  After some manipulation
\begin{equation}
\begin{split}
\Real z^2_k &= (\tilde x^R_k - \tilde y^R_k)^2
- (\tilde x^I_k - \tilde y^I_k)^2 
=
(x_k - y_k)^2 
+
\frac{1}{4}(v^R_k a_k - w^R_k c_k) \\
&+ (x_k - y_k)(v^R_k a_k - w^R c_k)
-\frac{1}{4}(a_kv^I_k - c_k w^I_k)^2.
\end{split}
\label{errorestimates:eqn2}
\end{equation}

Recall that $(x_k - y_k)^2 \geq \tau^2_{i,j}$ and suppose that there
is a positive constant $\delta_{k} > 0 $ such that
\begin{equation}
\delta_{k} \leq 
\frac{\sqrt{32\,\tau_{i,j} ^2+8\,\tau_{i,j} +1}-1 - 4\,\tau_{i,j}  }{4\,\tau_{i,j}}.
\label{errorestimates:eqn2a}
\end{equation}

\corb{Assume that $\lvert v^R_k \rvert\leq \tau_{i,j} \delta_{k} /
  a_{k}$, $\lvert v^I_k \rvert \leq \tau_{i,j} \delta_{k}/a_{k}$,
  $\lvert w^R_k \rvert \leq \tau_{i,j} \delta_{k} / c_{k}$, and
  $\lvert w^I_k \rvert \leq \tau_{i,j} \delta_{k} / c_{k}$.}  From
equations \eqref{errorestimates:eqn2} and \eqref{errorestimates:eqn2a}
it follows that
\begin{equation}
\Real z^2_k \geq \tau_{i,j}^2 (1 - 4 \delta_{k}^2) - \frac{\tau_{i,j}
  \delta_{k}}{2} > 0.
\label{errorestimates:eqn3}
\end{equation}
Furthermore,
\begin{equation}
\begin{split}
  \lvert \Real z^2_k \rvert &\leq 
(\max\{ \lvert y^{max}_k - x^{min}_{k} \rvert, \lvert x^{max}_k - y^{min}_{k} \rvert\})^2
\\
&+ \frac{1}{2}\tau_{i,j} \delta_{k} 
+ 
\max\{ \lvert y^{max}_k - x^{min}_{k} \rvert, \lvert x^{max}_k - y^{min}_{k} \rvert\} 
2\tau_{i.j} \delta_{k}
+ \tau^2_{i,j} \delta_{k}^2 \\
&\leq
1 + \frac{5}{2}\tau_{i,j} \delta_{k} + \tau^2_{i,j} \delta_{k}^2. 
\end{split}
\label{errorestimates:eqn6}
\end{equation}
Similarly,
\begin{equation}
  \lvert \Imag z^2_k \rvert = \lvert 2(\tilde x^R_k - \tilde y^R_k)(x^I_k - y^I_k) \rvert
  \leq
2 \tau_{i.j} \delta_{k} + 4 \tau^2_{i,j} \delta_{k}^2.
\label{errorestimates:eqn4}
\end{equation}

We now show how $\alpha_k$ and $\gamma_k$ can be extended into the
Bernstein ellipses $\mcE_{\sigma^{\alpha}_k}$ and
$\mcE_{\sigma^{\gamma}_k}$, for some $\sigma^{\alpha}_k > 0$ and
$\sigma^{\gamma}_k > 0$ such that $\Real z^2_k > 0$. \corb{Recall that
$\lvert v^R_k \rvert\leq \tau_{i,j} \delta_{k} /
  a_{k}$, $\lvert v^I_k \rvert \leq \tau_{i,j} \delta_{k}/a_{k}$,
  $\lvert w^R_k \rvert \leq \tau_{i,j} \delta_{k} / c_{k}$, and
  $\lvert w^I_k \rvert \leq \tau_{i,j} \delta_{k} / c_{k}$.}
These restrictions form a region in $\bbC \times \bbC$ and a Bernstein
ellipse is embedded (See Figure \ref{analyticity:fig1}).  This is done
by solving the following equation: $\frac{e^{\sigma^{\alpha}_k} +
  e^{-\sigma^{\alpha}_k}}{2} = 1 + \frac{\tau_{i,j}
  \delta_{k}}{a_k}$. The unique solution is
$\sigma^{\alpha}_k = \cosh^{-1} \left(1 +
\frac{\tau_{i,j} \delta_{k}}{a_k}
\right)$
with $\sigma^{\alpha}_k > 0$. Following a
similar argument we have that
$\sigma^{\gamma}_k = \cosh^{-1} \left(1
+ \frac{\tau_{i,j} \delta_{k}}{c_k}
\right)$
with $\sigma^{\gamma}_k > 0$. Let $\mcE^{d}_{\alpha} :=
\prod_{k=1}^{d} \mcE_{\sigma^{\alpha}_k}$ and $\mcE^{d}_{\gamma} :=
\prod_{k=1}^{d} \mcE_{\sigma^{\gamma}_k}$.  It follows that
\begin{equation}
\begin{split}
  \Real \vartheta &\geq  
  \sum_{k=1}^{d} \theta_k
  \Real z^2_k 
  \geq
\sum_{k=1}^{d} \theta_k 
  \left(
  \tau_{i,j}^2 (1 - 4 \delta_{k}^2) - \frac{\tau_{i,j}
    \delta_{k}}{2} \right)
  > 0.
  \end{split}
\label{errorestimates:eqn8}
\end{equation}
Thus there exist an analytic extension of $\phi(r;\btheta):\Gamma^d
\times \Gamma^d \rightarrow \R$ on $\mcE^{d}_{\alpha} \times
\mcE^{d}_{\gamma}$.

\begin{figure*}[htb!]
\begin{center}
\begin{tikzpicture}
    \begin{scope}[font=\scriptsize]

    \draw [->] (-2.5, 0) -- (2.5, 0) node [above left] {$\Real $};
    \draw [->] (0,-1.5) -- (0,1.5) node [below right] {$\Imag$};
    \draw [-,dashed] (-2,-1.5) -- (-2,1.5);
    \draw [-,very thick] (-1,0) -- (1,0);
    \draw [-,dashed] (2,-1.5) -- (2,1.5);
    \draw (1,-3pt) -- (1,3pt) node [above] {$-1$};
    \draw (-1,-3pt) -- (-1,3pt) node [above] {$1$};

    \draw [-,dashed] (-2,1) -- (2,1);
    \draw [-,dashed] (-2,-1) -- (2,-1);

    \fill [opacity=0.2, pattern=north west lines, pattern color=red]
    (-2,-1) rectangle (-1.5,1);

    \fill [opacity=0.2, pattern=north west lines, pattern color=red]
    (1.5,-1) rectangle (2,1);
    \end{scope}
    
    \node [below right] at (-1.92,0) {$\frac{\tau_{i,j} \delta_{k}}{a_k}$};
    \node [below right] at ( 1.05,0) {$\frac{\tau_{i,j} \delta_{k}}{a_k}$};

    \node at (0,-1.8) {$(a)$};
    \node at (6,-1.8) {$(b)$};

    \begin{scope}[shift={(0,0)},font=\scriptsize]

      \fill [pattern=north west lines, pattern color=red,semitransparent]
      (6,0) ellipse (2 and 1);

    \draw [->] (3, 0) -- (9, 0) node [above left] {$\Real $};
    \draw [->] (6,-1.5) -- (6,1.5) node [below right] {$\Imag$};
    \draw (5,-3pt) -- (5,3pt)   node [above] {$1$};
    \draw (7,-3pt) -- (7,3pt) node [above] {$-1$};

\end{scope}
    
    \node [below right] at (7.50,1.25) {$\mcE_{\sigma^{\alpha}_k}$};
    \node [below right] at (7.25,0.05) {$\frac{e^{\sigma^{\alpha}_k}
                                       + e^{-\sigma^{\alpha}_k}}{2}$};
        
\end{tikzpicture}
\end{center}
\caption{(a) Region of Complex extension of $\alpha_k$.  (b) Embedding
  of Bernstein ellipse $\mcE_{\sigma^{\alpha}_k}$.}
\label{analyticity:fig1}
\end{figure*}

The next step is to bound the absolute value of the Mat\'{e}rn
covariance function $ \lvert \phi(z;\btheta) \rvert$ in $\mcE^{d}_{\alpha} \times
\mcE^{d}_{\gamma}$. If $\nu > \frac{1}{2}$ and $\Real z>0$ then the
modified Bessel function of the second kind satisfies the following
identity:
$K_{\nu}(\sqrt{2 \nu}z) = \frac{\sqrt{\pi} (\sqrt{2 \nu}z)^{\nu}}{2^\nu
  \Gamma(\nu + \frac{1}{2})} 
\int_{1}^{\infty} (t^2 - 1)^{\nu -
  \frac{1}{2}} \exp{(-\sqrt{2 \nu}zt)}\, \text{d}t.$
It is not hard to show that for $\nu > \frac{1}{2}$ and $\Real z >0$,
we have that $ \lvert K_{\nu}(\sqrt{2 \nu}z) \rvert \leq \frac{ \lvert \sqrt{2
    \nu}z \rvert^{\nu}}{(\Real \sqrt{2 \nu}z)^{\nu}}$ $K_{\nu}(\sqrt{2 \nu}
\Real z)$.
Note that $r_{\vartheta} \geq \Real \vartheta > 0$.  From equation
\eqref{errorestimates:eqn4} we have that $\Imag \vartheta =
\sum_{k=1}^{d} \theta_k \Imag z^2_k \leq \sum_{k=1}^{d} 2 \tau
\delta_{k} + 4 \tau^2 \delta_{k}^2$.  From equation
\eqref{errorestimates:eqn8} $ \lvert \theta_{\vartheta} \rvert
  \leq
\xi(\btheta,\bdelta,\tau_{i,j}) $.

Since $K_{\nu}(\cdot)$ is strictly completely monotonic
\cite{Baricz2011} then
\begin{equation}
\begin{split}
   \lvert K_{\nu}(\sqrt{2 \nu}\Real z) \rvert &=
   \lvert K_{\nu}\ (\sqrt{2 \nu}
  r_{\vartheta} \cos(\theta_{\vartheta}/2))
  \rvert
  \leq
  \lvert K_{\nu}\Big(\sqrt{\frac{\nu}{2}}
  \cos(\xi(\btheta,\bdelta,\tau)/2) \\
  &
  \sum_{k=1}^{d} \theta_k 
  \Big(
  \tau_{i,j}^2 (1 - 4 \delta_{k}^2) - \frac{\tau_{i,j}
    \delta_{k}}{2} \Big)
  \Big) \rvert.
\end{split}
\label{errorestimates:eqn5}
\end{equation}
From equations \eqref{errorestimates:eqn6}
\eqref{errorestimates:eqn4} 
\begin{equation*}
\begin{split}
\lvert z_k \rvert^{2} &\leq \lvert \Real z^2_k \rvert + \lvert \Imag z^2_k \rvert
\leq
\mcR(\delta_k,\tau_{i,j}) :=
1 + \frac{9}{2}\tau_{i,j} \delta_{k} + 5 \tau^2_{i,j} \delta_{k}^2
\end{split}
\end{equation*}
and therefore
\begin{equation}
\begin{split}
  \lvert z \rvert
  &\leq
\lvert
\sum_{k=1}^{d} \theta_{k} z_k^{2} \rvert^{\frac{1}{2}}
\leq
\left(
\sum_{k=1}^{d} \theta_{k} \lvert z_k \rvert^{2} \right)^{\frac{1}{2}}
\leq \left( \sum_{k=1}^{d} \theta_k \mcR(\delta_k,\tau_{i,j})
\right)^{\frac{1}{2}}.
\end{split}
\label{errorestimates:eqn7}
\end{equation}
By combining equations \eqref{errorestimates:eqn4},
\eqref{errorestimates:eqn8}, \eqref{errorestimates:eqn5}, and
\eqref{errorestimates:eqn7}, we have now proven the Theorem.

\end{appendix}

\section*{Declarations}

\begin{itemize}
\item This material is based upon work supported by the National Science
  Foundation under Grant No. 1736392 and No. 2319011.
 \item  The authors declared that they have no conflict of interest.
\end{itemize}

\bibliographystyle{plain}
\bibliography{citations,multilevel}

\end{document}